\documentclass[11pt]{article}
\pdfoutput=1

\usepackage{jheppub}

\usepackage{amsmath}
\usepackage{amssymb}
\usepackage{amsfonts}
\usepackage{mathrsfs}

\usepackage{setspace}
\usepackage{bm}
\usepackage{bbm}

\usepackage[normalem]{ulem}
\usepackage{enumerate}
\usepackage{yfonts}

\usepackage{psfrag}

\usepackage[latin1]{inputenc}
\usepackage{graphicx}
\usepackage{cancel}
\usepackage{slashed}

\usepackage[font=small,labelfont=bf]{caption}
\usepackage{subcaption}

\usepackage[colorlinks=true]{hyperref} 
\hypersetup{
    bookmarks=true,         
    unicode=false,          
    pdftoolbar=true,        
    pdfmenubar=true,        
    pdffitwindow=false,     
    pdfstartview={FitH},    
    pdftitle={Instantons and Entanglement Entropy},    
    pdfauthor={Arpan Bhattacharyya, Ling-Yan Hung, and Charles M. Melby-Thompson},     
    pdfnewwindow=true,      
    colorlinks=true,        
    linkcolor=blue,         
    citecolor=red,          
    filecolor=magenta,      
    urlcolor=cyan           
}

\newcommand{\be}{\begin{equation}}
\newcommand{\ee}{\end{equation}}
\newcommand{\bes}{\begin{equation*}}
\newcommand{\ees}{\end{equation*}}
\newcommand{\bea}{\begin{eqnarray}}
\newcommand{\eea}{\end{eqnarray}}
\newcommand{\beas}{\begin{eqnarray*}}
\newcommand{\eeas}{\end{eqnarray*}}

\newcommand{\ba}        {\begin{align}}
\newcommand{\ea}        {\end{align}}

\newcommand{\req}[1]    {(\ref{#1})}

\newcommand{\p} {\partial}

\newcommand{\corr}[1]{\left\langle{#1}\right\rangle}

\newcommand{\ket}[1]{\left|{#1}\right\rangle}

\newcommand{\R} {\mathbb{R}}

\newcommand{\Z} {\mathbb{Z}}

\newcommand{\cD}    {\mathcal{D}}

\newcommand{\cR}    {\mathcal{R}}
\newcommand{\cS}    {\mathcal{S}}
\newcommand{\cZ}    {\mathcal{Z}}

\newcommand{\Tr}       {\mathrm{Tr}}

\newcommand\Group[1]    {\mathrm{#1}}
\newcommand\gO          {\Group{O}}
\newcommand\U           {\Group{U}}
\newcommand\SU          {\Group{SU}}

\begin{document}

\title{\bf Instantons and Entanglement Entropy}

\author[a]{Arpan Bhattacharyya,}
\author[a,b,c]{Ling-Yan Hung}
\author[a]{and Charles M. Melby--Thompson}

\affiliation[a]{Department of Physics and Center for Field Theory and Particle Physics,\\Fudan University, 220 Handan Road, 200433 Shanghai, China}
\affiliation[b]{State Key Laboratory of Surface Physics and Department of Physics,\\Fudan University, 220 Handan Road, 200433 Shanghai, China}
\affiliation[c]{Collaborative Innovation Center of Advanced  Microstructures,\\
Nanjing University, Nanjing, 210093, China.}

\abstract{We would like to put the area law --- believed to be obeyed by entanglement entropies in the ground state of a local field theory --- to scrutiny in the presence of non-perturbative effects. We study instanton corrections to entanglement entropy in various models whose instanton contributions are well understood, including $\U(1)$ gauge theory in 2+1 dimensions and false vacuum decay in $\phi^4$ theory, and we demonstrate that the area law is indeed obeyed in these models. We also perform numerical computations for toy wavefunctions mimicking the theta vacuum of the (1+1)-dimensional Schwinger model. Our results indicate that such superpositions exhibit no more violation of the area law than the logarithmic behavior of a single Fermi surface.
}

\maketitle
\flushbottom

\section{Introduction}
Recent breakthroughs in the study of many-body entanglement 
have led to a deeper understanding of phases of matter and their classification, driven in large part by the advances in the study of many-body entanglement.
This viewpoint was crucial in moving beyond the Landau-Ginzburg paradigm, meeting with tremendous success in classifying symmetry-protected topological (SPT) phases and topological orders.
It has also been used to help uncover entirely new phases, loosely known as quantum solids, and for the classification of quantum systems in~\cite{Swingle}, where the notion of ``s-source renormalization'' was introduced. S-source renormalization is a renormalization group (RG) scheme that focuses on entanglement. The key idea is to determine the excess entanglement that has to be incorporated into a many-body system to prepare the \emph{same} phase with the number of sites doubled. The parameter $s$ is the number of copies of the system at size $L$ needed in order to produce a system of size $2L$. Without going into details, suffice it to say that under this classification, matter belonging to different classes carry different amounts of entanglement. Matter whose entanglement entropy satisfies an area law is characterized by the value $s=1$. It is only one of the many possible classes of matter. 

The high-energy community, on the other hand, is predominantly concerned with continuum field theory.
This has long been the primary framework used to study fundamental particles and their interactions, and continuum field theory is also one of the most successful tools in the  study of condensed matter systems.
This naturally leads to the question: what phases of matter can continuum field theory describe?
Certainly, there are limits to what can be realized in field theory: for example, as is familiar to condensed matter physicists, not every lattice model has a continuum limit. 
An example is given by the Haah code \cite{Haah}, one of the ``quantum solids'' mentioned above, in which the jump in the ground state degeneracy between an even and an odd number of lattice sites is expected to preclude a continuum description.
It is therefore natural to assume that the phases of matter describable by local field theory are highly restricted.
In fact, it is argued in \cite{Swingle} that \emph{all} local quantum field theories belong to the class $s=1$, meaning that they must satisfy an area law. The argument purportedly works independently of whether the theory has a perturbative limit, and so it is important to subject this claim to scrutiny, particularly in regimes which have not received much attention from this perspective.

It is expected that, at any finite order in perturbation theory, the ground state entanglement receives only short-range corrections.
Motivated by these considerations, we initiate here a study of contributions to entanglement entropy coming from non-perturbative effects in field theory, especially within the instanton formalism.
As is well known, instantons can be used to capture tunneling effects, and therefore provide crucial information characterizing the ground state of a system, as well as the non-perturbative decay of a system from a false vacuum to its true ground state.
Instanton contributions are accompanied by an integral over the instanton moduli space, leading to volume contributions to the partition function.
Within the replica trick, the cancellation of the volume dependence between the partition function on the replica space, $Z_n$, and the $n^\text{th}$ power of the flat space partition function, $(Z_1)^n$, is required to recover the area law, making instantons a natural place to look for area law violations.
The aim of this note is to investigate the effect of instantons on entanglement entropy in several contexts.

We begin in section~\ref{sec:replica trick} by outlining the computation of instanton contributions to entanglement using the replica trick.
Probably the most popular tool for computing entanglement entropy in field theory, 
the replica trick introduces $n$ copies of the ground state wavefunction, sews them together in a non-trivial way, analytically continues in $n$, and finally takes the $n\to 1$ limit \cite{Callan0,Cardy}.
In the path integral formalism, this process boils down to computing the path-integral of the theory in a background with a conical singularity.

Our first application is to $\U(1)$ gauge theory in 2+1 dimensions, which has a dual description as the XY model.  
In the XY model, which is gapped, the entanglement across entangling surfaces much larger than the gap scale saturates to a simple area law.
In the gauge theory description, the gap arises due to non-perturbative effects, and so if we want the qualitative behavior of entanglement at large scales to match between the two descriptions, we must include contributions to the entanglement entropy arising from instanton configurations.
This system demonstrates the importance of instanton contributions for entanglement entropy, and provides a test of their computation via the replica trick.
In section~\ref{sec:vacuum decay} we turn to entanglement entropy of the false vacuum in $\phi^4$ theory in $1+1$ dimensions.
Upon integrating over classical solutions that break the replica rotation symmetry when the number of sheets $n$ is an integer, one recovers an area law correction to the entanglement entropy.
Section~\ref{sec:theta vacuum} examines a toy model for the vacuum of the Schwinger model in terms of a sum over Fermi surfaces.
The results of both these sections are consistent with conventional wisdom. 

We are thus led to the conclusion that non-perturbative corrections in field theory conform to the area scaling of ground states of local field theory. 
We further conclude that the time dependence of the entanglement entropy during non-perturbative vacuum decay obeys the Lieb-Robinson bound, \emph{i.e,} the rate of change of entropy is bounded by an area law.

Let us note that instantons have featured in the literature in the context of supersymmetric Renyi entropy, where supersymmetric localization can be applied. (See, for example, \cite{Huang, Nishioka}.) In those studies, their contributions either happen to cancel out, or are sufficiently localized near the conical singularity that no modification of the area law is observed. 
Due to the fact that supersymmetric Renyi entropy carries rather different physics, and that instabilities of the type considered here do not arise in supersymmetric theories, our study offers a useful complement to these works.

\section{Instantons, Entanglement, and the Replica Trick}
\label{sec:replica trick}
Instantons are saddle points of the Euclidean path integral. 
Their contribution is typically suppressed by a factor of the form $e^{-c/g}$, with $g$ some coupling constant of the theory.
For computations in quantum field theory, which are usually only under 
control at small coupling, these contributions are therefore much smaller than the perturbative corrections, if present.
As a result, for most computable quantities, instanton contributions are not quantitatively important.

There are a couple of important exceptions to this rule.
One is cases where some symmetry --- usual supersymmetry --- allows exact computations even at strong coupling.
Another important exception is when instantons lead to 
\emph{qualitative} changes in the behavior of physical quantities:
thus, instantons play a crucial role in processes such as non-perturbative 
vacuum decay, where the perturbative answer is \emph{zero}, and in the 
dynamics of 3d $\gO(3)$ gauge theory, where instanton condensation induces a gap.
Our primary interest, therefore, is in instanton corrections that lead to
qualitative changes in the entanglement entropy.
The three-dimensional $\gO(3)$ Higgs model (or lattice $\U(1)$ gauge theory) 
is a prime example of this, as instanton effects take a theory that is 
perturbatively gapless, and hence has entanglement entropy that depends 
non-trivially on system size, and gives it a gap, causing the entanglement
entropy to saturate at the scale of the gap.

Our goal in this note is to compute the entanglement entropy when the entanglement region is 
the half-space, $x_1>0$, using the replica trick. 
We proceed by replacing the $d$-dimensional Euclidean spacetime by the
product of a cone in the $(\tau,x_1)$ plane and $\R^{d-2}$.
The conical background with angular surplus $2\pi(n-1)$ has metric given by
\be
    ds^2_n  = dr^2 + n^2 r^2\,d\theta^2 + \sum_{i=2}^{d-1} dx_i^2, 
    \qquad 
    re^{i\theta}=x_1+i\tau\,,
    \quad
    \theta\sim\theta+2\pi\,.
\label{eq:replica metric}
\ee
Denote by $Z^{(n)}$ the path integral on the cone geometry.
When $n$ is an integer, the R\'enyi entropy is given by the formula
\be \label{eq1}
S_{n}=\frac{\log[Z^{(n)}]-n \log[Z^{(1)}]}{1-n}.
\ee
The entanglement entropy $S_{EE}$ can by computed by analytically continuing in $n$ and taking the limit $n\to 1$.
Expanding in a power series in $(n-1)$,
\be
\log[Z^{(n)}]=\log[Z^{(1)}]+(n-1)\frac{\partial}{\partial n} \log[Z^{(n)}]\Big\vert_{n=1}+\cdots \,,
\ee
gives
\be
S_{EE}=-\frac{\partial}{\partial n} \log[Z^{(n)}]\Big\vert_{n=1}+\log[Z^{(1)}].
\label{eq:EE}
\ee

In the semi-classical limit, the path-integral should include, in addition to 1-loop perturbative effects, a sum over classical saddles. 
These saddles are the instanton solutions.
The contribution from such instantons schematically takes the form
\be
\sum_I
\int d\mu_I(s)\, e^{-S_I} {\det}'\, \cD
\,.
\ee
Here $I$ labels instanton sectors, $s$ and $d\mu$ are the coordinates and measure on the instanton moduli space in sector $I$, and $\det'$ is the determinant with zero modes omitted.
For example, for a single instanton solution whose only moduli are the coordinates of its center $x$ the measure takes the form $d\mu=d^dx\,(S_1/2\pi)^{d/2}$, with $S_1$ the instanton action.

The moduli space arises when there are flat directions in the action, the simplest of which are due to translation invariance, and the measure arises from the integration over these flat directions -- the zero modes of $\cD$.
On the replica spacetime, translation invariance is broken, lifting the zero modes -- \emph{i.e.,} there are no saddles localized about generic points in Euclidean spacetime.
This does not however mean that such configurations should be dropped from the path integral: as instantons are separated from the conical singularity, they ``forget'' they are not on flat space and become arbitrarily close to saddle points.
The same situation arises in the presence of instanton--anti-instanton configurations: these only become saddles asymptotically at large separations.

In theories that are infrared-free and gapless (at least before instanton effects are included), instanton effects are essentially captured by long-range free field configurations. 
The measure for a single instanton in this case is captured explicitly by the long-distance interactions with other instantons and with itself, which will play a r\^{o}le in section~\ref{sec:gauge theory}.

The second sort of situation we will discuss involves well-localized instantons in theories with a gap.
In this case, the appropriate approach is to perform a sum over \emph{constrained saddles}.
In systems at weak coupling, for generic instanton configurations, the distance scale over which falloff occurs is of the order of coupling.
As a result, when expanding the constrained saddles in the perturbation expansion, the integration region in which the effect of interactions (either with the singularity or with other instantons) is important, is small.%
\footnote{In section~\ref{sec:vacuum decay} we will deal with situations where the instantons are large, but interactions are only important when two instanton walls are very close. The same considerations apply there.}
Moreover, the instanton action is large, which means that for generic configurations, the contributions to the measure coming from this source are much larger than that from such interactions.

The upshot is that at leading order in the perturbation theory expansion around instanton backgrounds, one can replace the small non-zero eigenvalue of $\cD$ by zero, and for the purposes of the instanton measure treat the integration over instanton centers as being flat.

\section{The \texorpdfstring{$\U(1)$}{U(1)} gauge theory in (2+1)-d as a pedagogical example}
\label{sec:gauge theory}
We begin our study of instanton effects with gauge theories. 
Here we consider two important roles played by instantons in gauge theory.
The first is in the structure of the ground state: while the field strength vanishes \emph{in vacuo}, there can be pure gauge configurations that nonetheless are classified by a discrete number, the homotopy class of the gauge potential at infinity.
To go from this infinite class of na\"ive vacua to the true vacuum requires a linear combination of such topological classes.
The result is called the theta vacuum.
The transition amplitude between different topological classes is captured by gauge instantons, which therefore play an important role in the structure of the true ground state of a gauge theory.

The second is confinement in 2+1 dimensions.
Here, a theory whose classical behavior in the infrared is that of free $\U(1)$ theory can be gapped by the inclusion of instanton configurations carrying a non-trivial first Chern class with respect to the infrared $\U(1)$.
As shown by Polyakov~\cite{Polyakov,Polyakov1}, while instantons are unlikely on small scales, at sufficiently large distances virtual instanton pairs proliferate and confinement results.

In both of these cases, it is reasonable to expect that instantons will play a role in the behavior of the entanglement entropy.
In this section we will focus primarily on the second, because it must have an important \emph{qualitative} effect on the entanglement entropy: when a theory has finite correlation length, the entanglement entropy is insensitive to the size of a sufficiently large system.
In this case therefore, instanton contributions play a decisive role in obtaining the correct behavior, even at weak coupling.
It is also a particularly nice case because we can compare directly to the dual description as an XY model.
We end the section with some comments on the theta vacuum.

\subsection{Free \texorpdfstring{$\U(1)$}{U(1)} gauge theory in \texorpdfstring{$d=3$}{d=3} and the XY model}
Our first stop is instanton contributions in $\U(1)$ gauge theory%
\footnote{As opposed to a theory with gauge group $\R$: compactness is important here.}
in 2+1 dimensions.
Since our interest is in instanton effects, we will work in Euclidean signature.
The Euclidean action is
\be
S = \int d^3 x\, \frac{1}{4e^2} F^{\mu\nu}F_{\mu\nu}
\ee
with equations of motion $\p_\mu F^{\mu\nu}=0$. 
It is useful to Poincar\'e dualize by writing 
$F_\mu = \frac{1}{2}\epsilon_{\mu\nu\sigma}F^{\nu\sigma}$,
so the equations of motion become
\be
\p^\mu F_\mu = \p_{[\mu}F_{\nu]} = 0 \,.
\ee

In flat space, instantons are the localized solutions of the form
\be
F_\mu(x) = -\frac{q}{2}\frac{x_\mu}{r^3}
\ee
with $q\in\Z$.
They have non-vanishing first Chern class $\int_{S^2}c_1(F)=q$ on $\R^3\setminus 0$.%
\footnote{In many applications --- for example, in the Abelian Higgs model studied by Polyakov --- $q$ is quantized in multiples of $2$ \cite{Polyakov,Polyakov1}.}
In pure $\U(1)$ gauge theory such configurations have infinite action, but in lattice gauge theory, or in the infrared limit of a non-Abelian gauge theory with maximally broken gauge group, such configurations describe the IR behavior of configurations with finite action.
Their action is of the order $\Lambda/e^2$, where $\Lambda$ is the scale at which the ultraviolet behavior becomes important.
The details of the UV completion are not important here, beyond the particular value of the instanton action.

These solutions have a simple interpretation in terms of the dual scalar field $\chi$ defined by $F_\mu = \p_\mu\chi$.
Compactness of the gauge group implies that $\chi$ is $2\pi$-periodic.
In this picture, the instanton background centered at $x_0$ is the response to a delta function source:
\be
\chi_{x_0}(x) = G(x,x_0) = \frac{1}{|x-x_0|}
\qquad\quad
\Box G(x,x_0) = -4\pi\delta(x-x_0)
\ee
(note the normalization of $G$).
These configurations arise as a natural component of the duality between $\U(1)$ gauge theory and the XY-model.
In the path integral, we must sum over general configurations with $N$ instantons with all allowed values of the charge $q$; the behavior of the theory is dominated however by the instantons with the smallest charge $q$ and their anti-instantons.

\subsection{Instantons in the replica spacetime}
\label{Ob}
Instanton solutions on the replica geometry are found the same way as in Euclidean space: requiring $\chi$ to be a classical saddle of non-trivial Chern class fixes it to take the form of the Green's function on the replica spacetime, $G^{(n)}$.
To evaluate the path integral, we should sum over multi-instanton configurations: 
\be
F^{(N;n)}_{\mu}(x) = \sum^N_{a=1} \frac{q_a}{2} \partial_\mu G^{(n)}(x;x_a)
\,,
\ee
with $q_a\in\Z$ the charge of the $a^\textrm{th}$ instanton and $x_a$ its position.
The action of this solution is
\begin{align}
S^{(N;n)} &= \frac{1}{2e^2}\int d^3x \sqrt{g_{(n)}}\, (F^{(N)}_{\mu})^2 
= \frac{1}{8e^2} \sum_{a,b} q_a q_b \int d^3x \sqrt{g_{(n)}}\, 
g_{(n)}^{\mu\nu}\p_\mu G^{(n)}(x;x_a)\p_\nu G^{(n)}(x;x_b) \nonumber \\
&= -\frac{1}{8e^2} \sum_{a,b} q_a q_b \int d^3x \sqrt{g_n}\, 
G^{(n)}(x;x_a) \Box_{(n)} G^{(n)}(x;x_b)
= \frac{\pi}{2e^2} \sum_{a,b} q_a q_b G^{(n)}(x_a;x_b)
\label{Ninstanton-action}
\end{align}
where we have dropped a boundary term in the second line.
The terms in this expression with $a=b$ are UV divergent, corresponding to the action of a single instanton.
In systems where the $\U(1)$ gauge field description only holds up to some
UV scale (\emph{e.g.} the Polyakov model or lattice gauge theory) this 
quantity is large but finite.
We continue to denote these terms formally by $G^{(n)}(x_a,x_a)$.
While formally infinite, they have an important (and finite) $n$-dependence
that we return to below.
In the special case where $n=1$, \req{Ninstanton-action} reduces to the standard 
result 
\be
S^{(N;1)} = \frac{\pi}{2e^2}\sum_{a\neq b} \frac{q_a q_b}{|x_a-x_b|} 
+ \sum_{a}q_a^2\cdot(\textrm{single instanton action})
\ee
The path integral follows from summing over all instanton solutions.
Contributions from instantons with composite charge are more highly suppressed, so that the leading correction involves only instantons of the smallest possible instanton charge $q$.
These take the form
\begin{align}
Z^{(n)} &\propto 
\sum_N \frac{1}{N!}\int\prod_{a=1}^N \left(d^3x_a\,\xi^{(n)}_a\right)
\sum_{q_a=\pm q}
    \,\exp\biggl(-\frac{\pi}{2e^2} \sum_{a\neq b} q_a q_b\, G^{(n)}(x_a;x_b)\biggr) 
\,, 
\label{Ninstanton-PF}
\end{align}
where we have absorbed the single instanton action, formally $G(x,x)$, into a measure term $\xi^{(n)}_a$.
The measure term, unlike in the flat case, does depend on the separation of the instanton from the replica singularity.

\subsubsection*{Comparison to the dual theory}
The dual description is by a $2\pi$-periodic scalar field $\chi$ with action
\be \label{dualaction}
S_{XY} = \frac{e^2}{4\pi^2} \int d^d x \left(\frac{1}{2}(\nabla\chi)^2 - M^2 \cos\chi\right)
\,.
\ee
Expression~\req{Ninstanton-PF} at $n=1$ is known to be recovered by expanding this
in powers of $M^2$~\cite{Polyakov1}. 
In our case, $d=3$ and $M^2=\frac{8\pi^2\xi^{(1)}}{e^2}$.
This is the famous duality between the compact $\U(1)$ gauge theory and the XY model.

From expression~\req{Ninstanton-PF}, we can see that the correspondence is unchanged for $n>1$:
the Green's function factor arises from the instanton interaction on one side of the duality, and from the 2-point function $\corr{e^{iq_a\chi}e^{iq_b\chi}}$ on the other.
This implies that the partition functions for these two effective field theories coincide on the replica spacetime, and hence that their entanglement entropies match.

We will discuss in the next section how these contributions can be computed on both sides of the duality, and make an explicit computation of $\p_n\cZ^{(n)}_N\vert_{n=1}$ in $(1+1)$ dimensions in section~\ref{sec:d=2}.

\subsection{Correction to the entanglement entropy}
Instanton contributions to the entanglement entropy arise from the terms $\p_nG^{(n)}\vert_{n=1}$.
Let us derive an expression for this quantity.
Consider the replica metric 
\be
ds^2_{n} = dr^2 + n^2r^2\,d\theta^2 + dx_2^2 
\ee
where $\theta\sim\theta+2\pi$; at $n=1$, the relation to the Euclidean coordinates $\tau=x^0$ and $x^1$ is $x_1+i\tau=re^{i\theta}$.
(The motivation for these coordinates is that neither the infrared cutoff nor the periodicity of $\theta$ depend on $n$.)
The Green's function at general $n$ satisfies the relation
\be
-\Box^{(n)}G^{(n)}(x;x') = 4\pi\delta_n(x,x')
= \frac{4\pi}{\sqrt{g_{(n)}}}\delta^{(3)}(x-x')
\label{gf-eq}
\ee
with $\delta_n(x,x')$ the covariant delta function and $\Box^{(n)}=\nabla_{(n)}^2$.
We define the following expansions:
\be
G^{(n)} = G^{[0]}+(n-1)G^{[1]} + \cdots \,,
\qquad\qquad
\sqrt{g_{(n)}}\nabla_n^2 
= \sqrt{\bar g}(\p^2 + (n-1)\Box^{[1]} + \cdots) 
\ee
where $\p^2$ denotes the flat space Laplacian, and $\bar g$ the flat space metric in a general coordinate system.
More precisely, 
$\Box^{[1]} = \p^2 - \frac{2}{r^2}\p_\theta^2 =: \p^2 - \cD^{[1]}$.
Multiplying both sides of \req{gf-eq} 
by $\sqrt{g_{(n)}}$ and expanding in $n$, we obtain the
relation
\be
    \p^2(G^{[0]}+G^{[1]}) = \cD^{[1]} G^{[0]} 
    \,,
\ee
and inverting $\p^2$ by $G^{[0]}$ gives the desired result \cite{subir}
\be
\p_n G^{(n)}(x;x')\vert_{n=1}
= G^{[1]}(x;x') = -G^{[0]}(x;x')
- \frac{1}{4\pi} \int d^3y\sqrt{\bar g}\, G^{[0]}(x;y)\, \cD_y^{[1]}G^{[0]}(y;x') \,.
\label{eq:perturbative method}
\ee

\smallskip
This expression is true for any mass, but when the classical field theory is a free CFT we can apply CFT methods. 
This corresponds to expanding the dual theory~\req{dualaction} around the conformal point $M^2=0$.
In this case~\cite{Smolkin}, the $O(n-1)$ correction to the Green's function $G^{(n)}(x;x')=\frac{e^2}{\pi}\corr{\chi(x)\chi(x')}_n$ in the replica spacetime is 
\be
G^{[1]}(x;x') - G^{[0]}(x;x') = \frac{e^2}{\pi} 2\pi\int_{\cR} d^{d-1}y\, y^1 \langle T_{00}(y) \chi(x) \chi(x')\rangle_c 
\,.
\label{eq:CFT G}
\ee
Here $\cR = \{\, y^\mu \,\vert\, y^1 > 0 \textrm{ and } y^0 = 0 \}$, the singularity is located at $y^1=y^0=0$, and $\langle \; \rangle_c$ is the connected CFT correlation function.
The three point function can be obtained using the methods of \cite{Osborn}, and for $d=3$ takes the form 
\be \label{TOO}
\corr{T_{00}(x_1) \chi(x_2)\chi(x_3)} 
= \frac{a\, h_{00}(\hat{X}_{23})}{|x_{12}|^3|x_{13}|^3|x_{23}|^{-2}}
\,.
\ee
The details are relegated to the appendix. 

As a simple and explicit illustration of this method, we calculate in the next section the contribution for the XY model in the simpler case of 1+1 dimensions in both duality frames.

\subsection{Explicit computation in 1+1 dimensions%
\label{sec:d=2}}
The low-energy effective action of the XY model in (1+1)d is simply that of a free real compact scalar,
\be
S_{XY} = \int d^2x\,  \frac{K}{2}  (\nabla\phi)^2 \,,
\ee 
with periodicity $\phi\simeq\phi+2\pi$.
This model has vortex solutions 
\be
\partial_\mu \phi(x) = -q \epsilon_{\mu\nu} \partial_\nu  \log |x-x_0|
\ee
where $q=\frac{1}{2\pi}\oint_{x_0} d\phi\in\Z$ is the vortex charge.%
\footnote{Sign conventions: $\epsilon_{12}=1$, and the integration contour is counter-clockwise.}
These are the $(1+1)$-d analogues of instanton solutions.
As in $(2+1)$-d, vortex configurations have infinite action in free continuum field theory, but an appropriate UV completion can render their action finite.
We assume this is the case.
Since the classical theory is free, a general $N$-vortex solution is given by the sum of vortices centered at $N$ different locations. 
As before, the multi-vortex configuration on flat space is written in terms of the two-dimensional Green's function of a massless scalar field,
\be
\p_\mu\phi(x) = -\epsilon_{\mu\nu}\p_\nu\sum_a \frac{q_a}{2} G(x;x_a) 
\ee
where 
\be
G(x;x') = - \log|x-x'|^2
\qquad
-\nabla^2 G(x;x') = 4\pi\delta^{(2)}(x-x') 
\,.
\ee
The vortices of the replica spacetime are obtained via the replacement $G\mapsto G^{(n)}$.
Computing as before, the $N$-vortex action on the replica geometry takes the form 
\be
S^{(N;n)} = \frac{1}{2}\pi K \sum_{a,b} q_a q_b  G^{(n)}(x_a;x_b) 
\ee
provided we require $\sum_{a}^N q_a = 0$.%
\footnote{The action of a single vortex also has an IR divergence, but configurations with total vortex charge zero are well-behaved.}
As before, the coincidence terms $\xi^{(n)}=G^{(n)}(x_a;x_a)$ of this expression are merely shorthand for the UV regulated values, but there are non-trivial IR contributions due to interaction with the conical singularity that are taken into account in what follows.
The contribution from the $N$-instanton sector now takes the form
\be
\cZ^{(n)}_N = \frac{1}{N!}\int\left(\prod_{a=1}^N d^2x_a\, \xi^{(n)}_a\right)
\sum_{\substack{q_a=\pm 1 \\ \sum q_a=0}} Z_N^{(n)}(\{q_a,x_a\}) 
\,,
\ee
\be
Z_N^{(n)}(\{q_a,x_a\}) = e^{-\frac{1}{2}\pi K\sum_{a\ne b}q_aq_bG^{(n)}(x_a;x_b)}
\,.
\ee

The entanglement entropy can be derived straightforwardly using the perturbative result~\req{eq:perturbative method}.
The relevant derivative is
\begin{multline}
\partial_n Z_N^{(n)} \bigr\vert_{n=1}
= \p_n \Bigl[\exp{-\frac{\pi}{2} K \sum_{a,b} q_a q_b G^{(n)}(x_a;x_b)} \Bigr]_{n=1} 
= -\frac{\pi}{2} K \sum_{a,b} q_a q_b G^{[1]}(x_a;x_b) \, Z_N^{(1)} ,
\end{multline}
where $G^{[1]}=\p_n G^{(n)}\bigr\vert_{n=1}$ as before.
Using complex coordinates $z=r e^{i\theta}$ we can write~\req{eq:perturbative method} in the form
\bea
G^{[1]}(x,x') 
&=& 
-G^{[0]}(x,x')
- \frac{1}{2\pi}
\int \frac{d^2z}{|z|^2}
    \bigl( z\p G - \bar z\bar\p G \bigr)
    \bigl( z\p G' - \bar z\bar\p G' \bigr) 
    \,,
\eea
where $G$ and $G'$ are shorthand for $G^{[0]}(z;x)$ and $G^{[0]}(z;x')$.%
\footnote{For $z=x+iy$, we take the measure $d^2z$ to mean $dx\,dy$.}
Integrating by parts and using $\Box G^{[0]} = -4\pi\delta$ gives
\be
G^{[1]} = 
+G^{[0]}
-\frac{1}{2\pi}
\int d^2z\Bigl( \frac{z}{\bar z}\p G\p G' + \frac{\bar z}{z}\bar\p G \bar\p G' \Bigr)
+ \frac{1}{2\pi}\oint G\vec\nabla G' \cdot d\vec n
\,.
\ee
Plugging in $G^{[0]}(x;x') = - \log |x-x'|^2$ and evaluating the integral for $|z| \le L$ ($L\gg |x|,|x'|$) gives a logarithmically divergent constant, together with a non-trivial finite contribution:
\be
G^{[1]}(x;x')-G^{[0]}(x;x') = -\log L^2 + \left(\frac{x\log x - x'\log x'}{x-x'} + \text{c.c.}\right)
+ O(L^{-1})
\,.
\label{eq:pert G}
\ee
(In this expression, $x$ and $x'$ denote complex numbers.) 
The divergent contribution drops out in the zero-charge sector.

In terms of the flat space instanton partition function 
\be
\cZ^{(1)} = \sum_{N=0}^\infty \frac{\xi^N}{N!}\left(\prod_{a=1}^N \int d^2x_a\right)
e^{-\frac{1}{2}\pi K\sum_{a\ne b}G^{(1)}(x_a;x_b)}
\ee
where $\xi=\xi^{(1)}$,
the instanton contribution to the entanglement entropy~\req{eq:EE} may be written
\begin{multline}
S_\text{EE, inst} = \log\cZ^{(1)} \\
+ \frac{1}{\cZ^{(1)}}\sum_{N=0}^\infty \frac{\xi^N}{N!}
\sum_{\substack{q_a=\pm 1 \\ \sum q_a=0}}
\left(\prod_{a=1}^N \int d^2x_a\right) 
\left(\prod_{a<b}|x_a-x_b|^{2\pi K q_a q_b}\right)
\frac{\pi K}{2} \sum_{a,b} q_a q_b G^{[1]}(x_a;x_b) \,,
\end{multline}
which can in principle be evaluated using~\req{eq:pert G}.
In this formula it is important that one interpret the coincidence terms $G^{[1]}(x;x)$ as
\be
\lim_{x'\to x}
G^{[1]}(x';x) = \log x + (\text{constant}) \,,
\ee
which represents the contribution of the conical surplus to the measure.

\medskip
We can check this perturbative computation by comparing with the dual theory, whose expansion in $M^2$ can be computed using CFT methods.
The dual field $\chi$ of periodicity $2\pi$ has, at $M^2=0$, the action
\be
S(\chi) = \frac{1}{8\pi^2K}\int d^2x\, (\p\chi)^2
\,,
\ee
allowing us to write
\be
Z_N^{(n)}(\{q_a,x_a\}) = e^{-\frac{1}{2}\pi K\sum_{a,b}q_aq_bG^{(n)}(x_a;x_b)}
= \corr{\prod_{a}e^{iq_a\chi(x_a)}} 
\ee
since $\corr{\chi(x)\chi(x')}=\pi K G(x;x')$. 
The stress tensor is
$T_{zz}=-\frac{1}{4\pi^2K}\!:\!\!\p\chi\p\chi\!\!:\,$.%
\footnote{This is the standard field theory stress tensor, $T^{\mu\nu}=\frac{2}{\sqrt{g}}\frac{\delta S}{\delta g_{\mu\nu}}$.}
Plugging $T_{00}=T_{zz}+T_{\bar z\bar z}$ into the CFT expression~\req{eq:CFT G} for $G^{[1]}$ gives
\begin{align}
\begin{split}
G^{[1]}(x;x') - G^{[0]}(x;x')
&= \frac{2\pi}{\pi K}\int_{0}^{\infty}dz\,z \Big( \corr{T_{z z}(z)\chi(x)\chi(x')} +\corr{T_{\bar z \bar z}(z)\chi(x)\chi(x')} \Big) \\
& = -\int_{0}^{\infty} dz\, z \Big(\frac{1}{(x-z)(x'-z)} + \frac{1}{(\bar{x}-\bar{z})(\bar{x}'-\bar{z})}\Big)  \,.
\end{split}\label{2dTOO}
\end{align}
Evaluating the integral, we find that it matches the perturbative result~\req{eq:pert G}.

Therefore we see that, once the instantons for the replicated geometry are known, expanding their contribution in $(n-1)$ recovers the correct contributions to the entanglement entropy.
Needless to say, here they obey the area law.

\subsection{Comments on SU(2) instantons in 3+1 dimensions}
Having discussed the simplest cases, let us offer some comments on non-abelian gauge instantons.
The simplest instanton contributions to $\SU(2)$ theories are the BPST instantons~\cite{BPST},
which satisfy the self-duality condition
\be
G = \tilde{G},
\ee
where $G = dA + A\wedge A$ is the $\SU(2)$ field strength and $\tilde{G}$ its Hodge dual. 
't~Hooft \cite{'tHooft} gave a simple ansatz for a family of flat space multi-instanton solutions in a singular gauge
\be
A^a_\mu = -\frac{1}{g} \tilde\eta_{a\mu\nu} \partial_\nu \log W(x), \qquad W = 1 + \sum_{i=1}^N \rho_i^2 \phi(x;x_i)
\qquad
\phi(x;x_i) = \frac{1}{(x-x_i)^2}
\,,
\ee
where $\rho_i$, which sets the instanton size, and the instanton centers $x_i$ are moduli to be integrated over in the path integral.
$\tilde\eta_{a\mu\nu}$ is the 't Hooft symbol 
\be
    \tilde\eta_{aij} = \epsilon_{aij}
    \qquad
    \tilde\eta_{a4i} = \delta_{ai}
    \qquad
    \tilde\eta_{ai4} = -\delta_{ai}
\,.
\ee

$W$ obeys two important properties: the multi-instanton solution is additive, and each term $\phi(x;x_i)$ in the sum is a Green's function for the 4-dimensional Laplacian.
Being harmonic away from the instanton centers guarantees $W$ is self-dual, while the particular property of the singularities allows them to be eliminated by a singular gauge transformation.
This suggests a simple way to generalize such solutions to an $n$-sheeted background: replace $\phi(x;x_i)$ by the Green's function of the replicated space
\be
    W(x) = 1 + \sum_i \rho_i^2 G^{(n)}(x;x_i)
\,.
\ee

General considerations imply that the action of these particular solutions is unchanged from flat space, as the action of a self-dual configuration must be quantized in units of $\frac{8\pi^2}{g^2}$;
the moduli space is modified, but only in such a way as to compensate for the increased volume of the replica spacetime.
Any non-trivial contribution from these instantons must therefore come from the one loop determinant, the computation of which is beyond the scope of the present paper.
It is also worth noting that non-self-dual configurations play a role in the path integral.
Since the action of such configurations is not quantized, it is natural to expect that in the replica spacetime, such configurations will give non-perturbative contributions to the entanglement entropy.

Finally, we note that gauge theory instantons in 4d are particularly interesting as regards entanglement entropy, since the existence of a size modulus leads to contributions from instantons of arbitrarily large spatial extent.
It is conceivable that the effects of the conical singularity could therefore be felt far away from the singularity, leading to a violation of the area law.%
\footnote{We would like to thank the referee for comments on this point.}
We leave a detailed analysis of 4d gauge instantons to future work.

\section{Entanglement entropy and false vacuum decay}
\label{sec:vacuum decay}
One of the quintessential applications of instanton methods is nonperturbative vacuum decay.
The simplest example is given by $\phi^4$ theory. 
Consider a scalar theory with Euclidean action~\cite{Coleman1,Coleman2,Coleman3}
\be
    L = \frac{1}{2} (\partial\phi)^2 +U(\phi),
    \qquad\qquad
    U = \frac{\lambda}{8} (\phi^2-a^2)^2 + \frac{\epsilon}{2a}(\phi-a),
\ee
where $\epsilon$ is a small positive number breaking the symmetry of the two minima located at $\phi = \pm a$. 
We choose the above $U(\phi)$ for concreteness, but any $U$ with two nearly-degenerate minima will do.
The dominant instanton configuration is radially symmetric around the instanton center, and satisfies the equation of motion
\be
    \partial_r^2 \phi(r) + \frac{(d-1)}{r}\partial_r\phi(r)  = \frac{dU(\phi)}{d\phi}
    \,.
\ee
This equation is difficult to solve in general, but there is a useful approximate solution if the single derivative term can be neglected. 
This is true if the thickness of the interpolation region is much smaller than the size of the instanton, which is known as the thin wall approximation.
In this approximation, the result for the particular potential above yields
\be
    \phi = a \tanh \bigl({ \tfrac{1}{2}}\mu (r-R) \bigr)
    \,,
\ee
where $\mu=a\sqrt{\lambda}$ is the effective mass around the minima $\phi=\pm a$.
Here the integration constant $R$ gives the instanton radius, found by extremizing the Euclidean action
\be
    S_E = {\rm vol}\,S^{d-1}\int dr\, r^{d-1}\left( \frac{1}{2}(\p_r\phi)^2 + U(\phi) \right)
    \approx T_{d-1} A - T_d V
\ee
where $T_{d-1} = \int_{-a}^a d\phi \sqrt{2U_0(\phi)}$ is the domain wall tension, $T_d = \epsilon$ is the difference in energy density between the two vacua, and $A$ and $V$ are the instanton surface area and volume.
Extremizing with respect to $R$ gives 
\be
    R = (d-1)\frac{T_{d-1}}{T_d}
\label{eq:thin wall R}
\ee
and thus also the action of a single instanton
\be
    S_0 = \frac{1}{d}T_{d-1}A 
    = \frac{1}{d-1}T_dV
    = \text{vol}\, S^{d-1}\,\frac{(d-1)^{d-1}T_{d-1}^d}{d\,\epsilon^{d-1}}
    \,.
\ee
Note that~\req{eq:thin wall R} implies the radius is arbitrarily large at small $\epsilon$, guaranteeing self-consistency of the approximation.

\subsection{Instantons in the dilute gas approximation}
The instanton solution of the $\phi^4$ theory describing tunnelling between a false vacuum and a true vacuum reviewed above is a typical example in which the single instanton solution is a localized field configuration falling off exponentially at large distances.
At weak coupling, the dominant contribution to the path integral comes from configurations whose instanton density is of order $e^{-S_0}$.
In this case, the contributing multi-instanton configurations are approximated well by the superposition of single instantons whose separation is much larger than their size. This is the dilute gas approximation. 

Leading contributions to the path integral come from three sources: the measure, the instanton action, and the 1-loop determinant.
The measure is determined by the instanton action through the presence of zero modes.
While exact zero modes are absent in multi-instanton solutions (as well as single instanton solutions in the replica spacetime), for small coupling these corrections can be ignored.
The final expression takes the form
\be
\cZ = \cZ^{(0)}\sum_{N}e^{-NS_0}\prod_{a=1}^N\left[ \frac{1}{2}\int d^dx_a \left(\frac{S_0}{2\pi}\right)^{d/2}\left(\frac{{\det}'\cD^{(1)}}{\det\cD^{(0)}}\right)^{-1/2} \right]
= \cZ^{(0)}e^{VKe^{-S_0}}
\ee
where $K=\frac{1}{2}(\frac{S_0}{2\pi})^{d/2}({\det}'\cD^{(1)}/\det\cD^{(0)})^{-1/2}$, and $V$ is the volume of Euclidean spacetime.
The factor of $\frac{1}{2}$ comes as usual from carrying out the correct analytic continuation of the path integral when $\cD^{(1)}$ has negative eigenvalues~\cite{Coleman2}.

\subsection{Single instanton contributions in \texorpdfstring{$d=2$}{d=2}%
\label{sec:domain wall instantons}}
We now turn to the computation of single instanton contributions to the entanglement entropy.
For simplicity we work in $1+1$ dimensions; although this case is special in certain particulars, we expect that the essential process is not modified in higher dimensions.

\begin{figure}
\begin{center}
\includegraphics[width=0.25\textwidth]{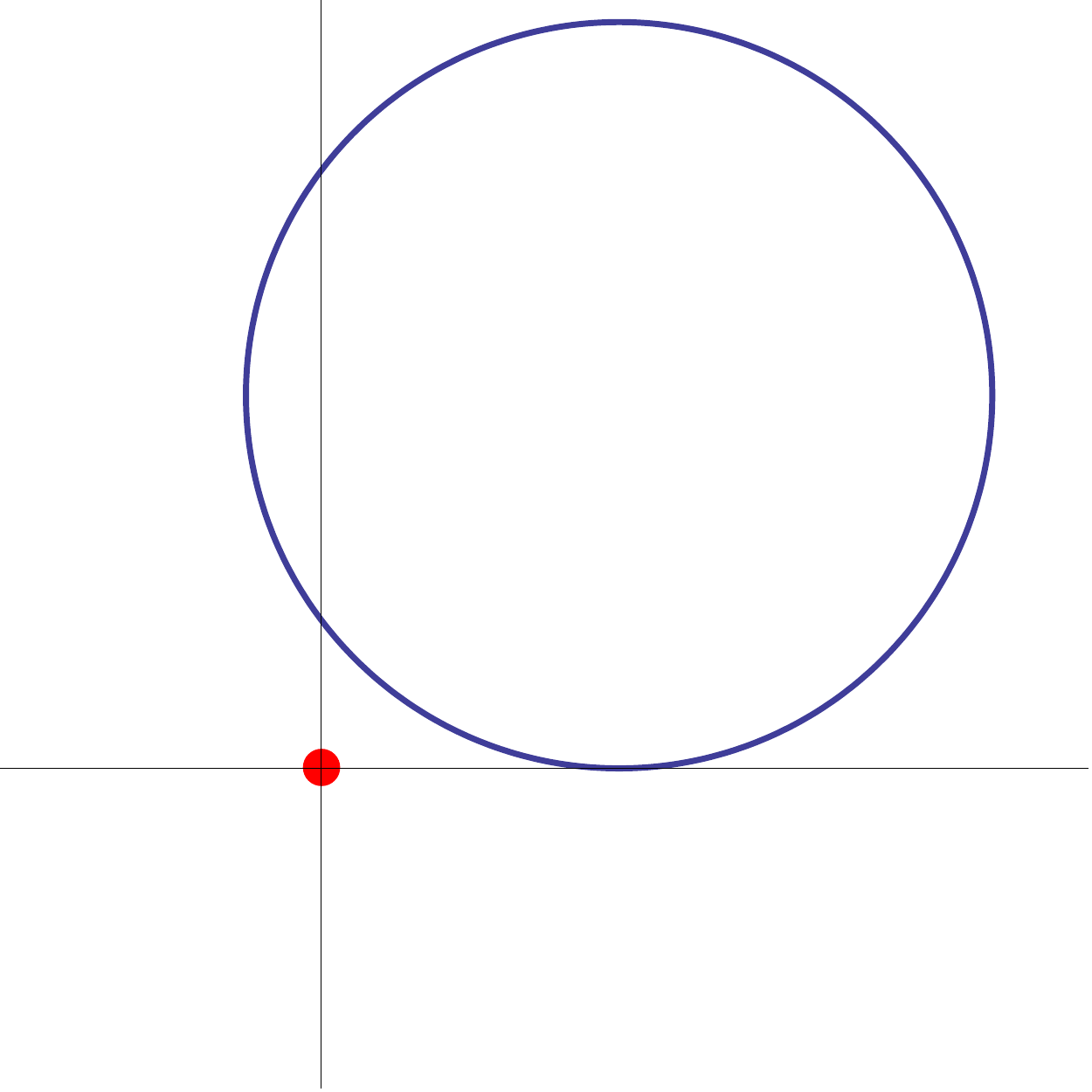}
\hspace{5ex}
\includegraphics[width=0.25\textwidth]{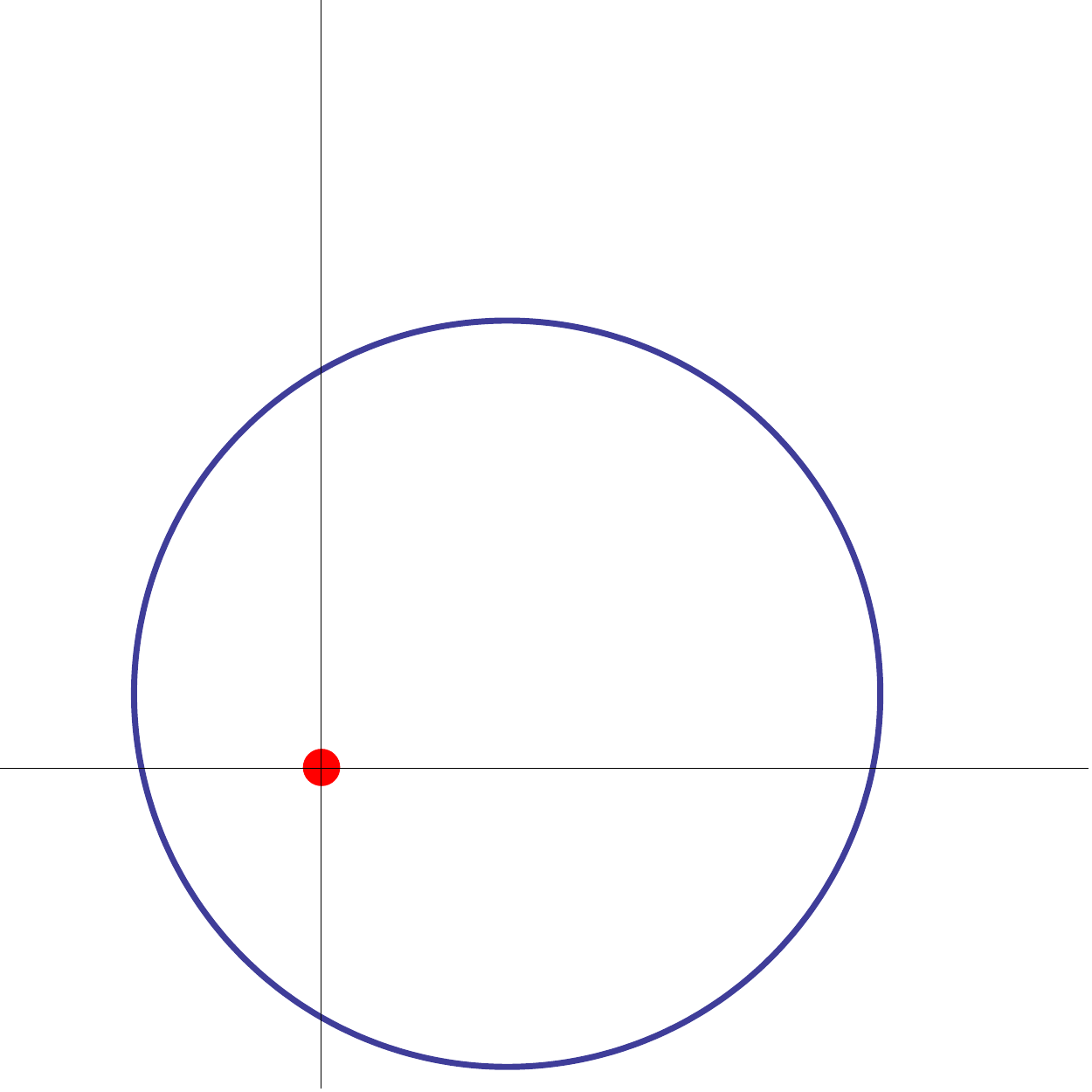}
\caption{The two types of instanton configurations. 
Left: The conical singularity lies outside the instanton configuration;
the instanton lies on a single sheet of the replica spacetime.
Right: Here the conical singularity lies inside the instanton wall.
The branch cut passes through the instanton wall
and the instanton extends over all $n$ sheets of the replica geometry.
}
\label{instanton-fig}
\end{center}
\end{figure}

Remember that the thin wall approximation implies that the instanton size is 
much larger than the inverse mass in the false vacuum.
This allows us to reduce the computation to two cases, shown in 
figure~\ref{instanton-fig}.

When the instanton is well-separated from the replica singularity (left side
of figure~\ref{instanton-fig}), the
solution is the same as in the flat spacetime; since $\phi$ is effectively
constant outside the bubble, there is no incompatibility with the $2\pi n$
periodicity of $\theta$ on the replica spacetime.
(In a single instanton configuration, of course, the instanton lives on
one replica sheet only.)
``Well-separated'' means that the separation of the wall from the singularity
is much larger than the wall thickness, but as the wall is thin compared to
its size, this is true for typical instanton configurations.
The action of such a configuration takes the same value $S_0$ as in flat 
spacetime.

To obtain the partition function we must also compute the relevant determinant and measure factors.
According to the discussion of section~\ref{sec:replica trick}, the complete expression is
\be
\frac{1}{2}
\int_{R}^\infty dr \, r\int_0^{2\pi n}d\theta\, 
\frac{S_0}{2\pi}e^{-S_0} ({\det}'{\cD^{(1)}_n})^{-1/2}
\ee
with $\cD^{(k)}_n$ the kinetic operator around the $k$-instanton background in the $n$-fold replica spacetime. 
($\det'$ is the determinant with zero modes omitted, and the $\frac{1}{2}$ comes from the presence of a single negative eigenvalue.)
Locality guarantees that to leading order the effects of the replica geometry and the instanton on the determinant are independent. 

A gross approximation to the determinant would ignore the instanton entirely, which would simply give a factor $({\det}'{\cD^{(0)}_n})^{-1/2}$.
This is of course modified by the presence of the instanton. 
As the instanton is located far away from the conical singularity, modes near the instanton wall will not see the conical singularity at all. 
Therefore,
$\frac{1}{2}\frac{S_0}{2\pi}(\det' \cD_n^{(1)})^{-1/2}=K\,(\det \cD_n^{(0)})^{-1/2}$
(plus subleading corrections). 
Here,
$K=\frac{1}{2}\frac{S_0}{2\pi}(\frac{\det'\cD_1^{(1)}}{\det\cD_1^{(0)}})^{-1/2}$
is the factor that arose in the false vacuum decay rate; writing things in terms of $K$ makes a comparison to the flat space partition function simple.
The final result is
\be
Z_n^{(0)}\cdot n \bigl(A-A(R)\bigr)Ke^{-S_0} + \mathrm{(subleading)}
\ee
with $A$ the total area of the flat Euclidean geometry (not the replica geometry), $A(R)=\pi R^2$ is the instanton area, and $Z^{(0)}_n$ is the 0-instanton partition function 
on the replica geometry.

When the singularity is contained well-inside the instanton bubble, as in
figure~\ref{instanton-fig}(b), the constancy of $\phi$ in the interior
once again implies that the
instanton solution has the same functional form; the difference is that
the cut intersects the instanton wall, and therefore the instanton extends 
to all $n$ sheets of the replica geometry.
The radius of curvature of the instanton wall is fixed by the domain wall 
tension, so that the classical solution will look like $n$ copies of the
same circle of fixed radius, each one glued to the next across the cut.
The action of the configuration is the length of the wall times its tension,
plus $V(\phi_-)$ times the area of the interior.
Both of these quantities are $n$ times the flat space value, and so the
action is simply $S=nS_0$.

In the measure factor, $S_0/2\pi$ is replaced by $nS_0/2\pi$;
on the other hand, because the instanton extends over $n$ sheets rather than
one, the solution is actually $2\pi$-periodic, and therefore we should only
integrate the angular coordinate of the instanton zero-mode over the interval
$(0,2\pi)$.
The contribution to $Z$ is therefore
\be
\frac{1}{2^{2n-1}}
\int_0^R dr\,r\int_0^{2\pi} d\theta\, \frac{nS_0}{2\pi}
({\det}'\cD^{<}_n)^{-1/2}e^{-nS_0} \,.
\ee
with $\cD^<_n$ the kinetic operator for $r<R$.
The factor $2^{1-2n}$ is included because we now have $2n-1$ negative eigenvalues (which we will confirm momentarily).

We must now evaluate the determinant.
It gets contributions from two widely separated scales.
On the one hand there are the ``fast'' modes associated to wave-like fluctuations 
of the scalar field; the eigenvalues of the kinetic operator on these modes
are $\gtrsim \mu^2$, and are the only ones present in the zero instanton sector.
They are responsible for the saturation of the entanglement
entropy at lengths $L\gtrsim m^{-1}$.
On the other hand, for the instanton background there is a new class of
``slow'' fluctuations associated to deformations of the domain wall.
Heuristically, we split the determinant into a product of contributions
from each class of modes, $\det'\cD=\det'_s\cD\cdot\det_f\cD$.
While strictly speaking the high-lying slow modes and low-lying fast modes
overlap, the high-lying slow modes are insensitive to the details of the
shape of the instanton, so that the factorization is accurate provided the
answer is written in terms of ratios of slow and fast mode determinants, such 
that the relevant operators have the same density of eigenvalues at 
scales much larger than $m^{-1}$ but much smaller than $R$.

The slow modes for the instanton come from the fluctuations of the domain
wall. 
If we allow the domain wall to fluctuate around a constant radius,
$r=R+\delta r(\theta)$, the corresponding fluctuation action is
\be
S^{(2)} = \frac{T_2}{2}\int_0^\beta d\theta\left((\delta r')^2 +
\lambda(\delta r)^2\right)\,.
\ee
For the instanton background, $\lambda=-1$, but we leave it as a free
parameter for future convenience.
The integration limit $\beta$ depends on the background: 
if the instanton encircles the origin then $\beta=2\pi n$, otherwise $\beta=2\pi$.
The slow contribution to the path integral is therefore given by
\be
\int D[\delta r]e^{-S^{(2)}(\delta r)}
\ee
The result, $[\det_s\cD]^{-1/2}$, is simply the harmonic oscillator partition function, and takes the value
\be
Z(\beta,\lambda) = \frac{1}{2\sinh(\beta\sqrt\lambda/2)}\,.
\ee

We are interested in the behavior of $Z(2\pi n,\lambda)$ as $\lambda\to-1$.
Note that $Z$ has a singularity at this point, due to the two zero modes 
associated to translations.
Since their contribution is already captured in the instanton measure, the corresponding eigenvalues should be omitted.
This is done via multiplication by $\left(\frac{2\pi n}{\beta}\right)^2+\lambda$.
(Note further that the $2n-1$ eigenmodes with eigenvalues
$\left(\frac{2\pi k}{\beta}\right)^2+\lambda$ for $0\le k < n$ are negative at $\beta=2\pi n$, $\lambda=-1$; 
all eigenvalues except for $k=0$ are two-fold degenerate.
These are the negative eigenvalues claimed above.)

Because it is doubly degenerate, we are left with
\be
Z'_n(\beta,\lambda) = \frac{(2\pi n/\beta)^2+\lambda}{2\sinh(\beta\sqrt\lambda/2)}.
\ee

We wish to compare with the slow modes in the single instanton sector,
in such a way that the eigenvalue density matches at high energies.%
\footnote{Appendix A of~\cite{Metlitski:2011pr} gives a computation that exhibits some similar features.} 
This is accomplished by comparing to $[Z'_1(2\pi,\lambda)]^n$, with
\be
Z'_1(\beta,\lambda) = \frac{(2\pi/\beta)^2+\lambda}{2\sinh(\beta\sqrt\lambda/2)}.
\ee
The final result is
\be
\left(\frac{\det_s'\cD^<_n}{[\det'_s\cD^{(1)}_1]^n}\right)^{-1/2}
= \lim_{\lambda\to -1}\frac{Z'_n(2\pi n,\lambda)}{[Z'_1(2\pi,\lambda)]^n} 
= \frac{(-i\pi)^{n-1}}{n}\,.
\ee

The ratio of the fast determinants can be argued to take the following form.
At distances from the domain wall well exceeding the correlation length, the path integral reduces to $n$ massive scalars, giving a vacuum determinant contribution $[\det_f\cD^{(0)}_1]^{-n/2}$. 
For a generic instanton contribution the domain wall is well-separated from the singularity, so that the contribution from modes localized near the domain wall can be taken into account by including a correction factor $[(\det_f\cD^{(1)}_1)/(\det_f\cD^{(0)}_1)]^{-n/2}$. 
Finally, we must take into account the behavior near the singularity by including a factor $[(\det_f \tilde\cD_n)/(\det_f  \tilde\cD_1)^n]^{-1/2}$; here $\tilde\cD$ is the kinetic operator expanded around the \emph{true} vacuum, whose mass generically differs from that of the false vacuum.

Putting everything together gives:
\be
{\det}_f(\cD_n^<) = \left(\frac{\det_f\cD_1^{(1)}}{\det\tilde\cD^{(0)}_1}\right)^n
\det\tilde\cD_n^{(0)}
\ee
(the bare replica geometry has no slow modes).
Therefore we obtain the contribution of a single instanton with $r<R$:
\begin{multline}
\frac{1}{2^{2n-1}}
Z_n^{(0)} \times
A(R)\frac{n S_0}{2\pi}e^{-nS_0} \left(\frac{\det_s' \cD_1^{(1)}\det_f \cD_1^{(1)}}{\det  \tilde\cD^{(0)}_1}\right)^{-n/2} 
\left(\frac{\det \tilde\cD^{(0)}_n}{\det \cD^{(0)}_n}\right)^{-1/2}
\frac{(-i\pi)^{n-1}}{n}\, \nonumber \\
= Z_n^{(0)} A(R)K e^{-nS_0} P_n{(-i\pi \kappa )^{n-1}}
\end{multline}
with $\kappa=\frac{1}{4}\left(\det'\cD_1^{(1)}/\det\cD_1^{(0)}\right)^{-1/2}$ and
\be
P_n = \left[
    \left(
        \frac{\det \cD_1^{(0)}}{\det \tilde\cD_1^{(0)}}
    \right)^n
    \frac{\det \tilde\cD_n^{(0)}}
        {\det \cD^{(0)}_n}
\right]^{-1/2}
\ee
(Recall $\kappa$ is imaginary because $\cD_1^{(1)}$ has a single negative eigenvalue.)

There is one further subtlety that arises in the case $d=2$, for here it is only possible for at most one instanton to wrap the conical singularity. 
The full partition function is therefore given by
\be
\mathcal{Z}_n = Z_n^{(0)} e^{n[A-A(R)]Ke^{-S_0}} \Bigl(1 + A(R)K e^{-nS_0}  P_n {(-i\pi \kappa )^{n-1}}\Bigr) 
\ee 
and so the entanglement entropy becomes
\be
\cS = S_1^{(0)}  +
 A(R)K P_1 e^{-S_0}(1 - (\log P_n)'\vert_{n=1} - \log(-i \pi \kappa e^{-S_0})).
\label{eq:instanton entanglement}
\ee
We note that the contribution of $P_n$ only depends on the relative entanglement entropy in the two vacua, and that in the special case where both vacua have the same mass, $P_n=1$.

There are several comments in order here.
Since the leading order instanton corrections to the entanglement entropy come entirely from instantons wrapping the singularity, in the limit of large intervals ($L\gg R$) the entanglement entropy will be insensitive to size, and hence in this limit the entanglement entropy satisfies the area law.
(We will come back to this point momentarily.)
More surprisingly, we should note as $K$ is imaginary \cite{Coleman1,Coleman2,Coleman3}, the non-perturbative correction to the entanglement is imaginary.
This raises the question of how one should interpret an imaginary entanglement entropy. 
Let us compare this computation with the original calculation in which the decay rate is extracted. 
There, one could understand the decay of the false vacuum wavefunction in time by Wick rotating the instanton result, in which
\be
\exp(i e^{-S_0} |K| V_{d-1} T ) \to \exp(- e^{-S_0} |K| V_{d-1} t ), \qquad T\to i t
\ee
Now, the area term $A(R)$ is a Euclidean volume term evaluated in the $x-T$ Euclidean plane, and so under Wick rotation $T \to i t$ its measure should acquire an overall factor of $i$.  
Therefore, including the contribution of $K$ recovers a real value. 

What is the meaning of this value? 
Our methods implicitly assume that the cutoff on Euclidean time is large compared to the instanton scale, and our result is therefore most naturally interpreted as the contribution after a time has passed that is much larger than the instanton size $R$ (in units where $c=1$), but small enough that the vacuum remains mostly undecayed.
Since our formulae really reflect the decay from the free false vacuum to the real vacuum, this suggests that there is an increase in entanglement entropy time scales $t\lesssim R$ that saturates for $t\sim R$, and whose final value is captured by our formula; while for $R\ll t\ll \Gamma^{-1/2}$ ($\Gamma$ is the vacuum decay rate per unit length), there is no time dependence.
On the other hand, we expect a non-trivial time dependence at subleading order in the perturbation theory expansion, which will pick up contributions from exponential tails in the domain wall shape, and from multi-instanton configurations.
The late-time dynamics of the entanglement entropy should presumably be picked up by these subleading contributions.
This suggestion seems consistent with the observations of \cite{QM} regarding tunneling in a two-particle quantum mechanics model, where the leading time dependence of the entanglement entropy appears to arise at two-instanton order. 

Of course, this is by no means a proof of how the formula should be interpreted, but only a guess suggested by  analytic continuation. 
We leave a more rigorous analysis to future work.

\subsubsection*{Entanglement entropy of a finite interval}
Above we evaluated the leading contribution to the entanglement entropy when the system is divided into two half-spaces.
Let us now turn to the entropy of entanglement between a finite interval of length $L$ and its complement.

The behavior of this case depends qualitatively on the relative size of $L$ and $R$.
When $R<L$, the computation is essentially above, except that the correction to the entropy (at leading order in $e^{-S_0}$) is twice as large, because we now have contributions from instantons centered at both endpoints of the interval. 
On the other hand, when $L<R$, there is an additional type of contribution, coming from those instantons which encircle the entire interval.
We divide the (approximate) moduli space of single instantons into three regions, $\cR_{0,1,2}$, labeling the number of singularities lying inside the true vacuum region.

In the new configuration, the instanton wall doesn't cross the branch cut, and therefore each connected component of the wall has length $2\pi R$.
However, since the branch cut lies in the true vacuum region, each sheet must have its own instanton wall.
Therefore the instanton action is $S=nS_0$.
Instantons on separate sheets however are free to move independently.
We must therefore integrate over $n$ instanton center of mass variables $x_n$, each with measure $\frac{S_0}{2\pi}d^2x_n$.

\begin{figure}
\begin{center}
\includegraphics[width=0.3\textwidth]{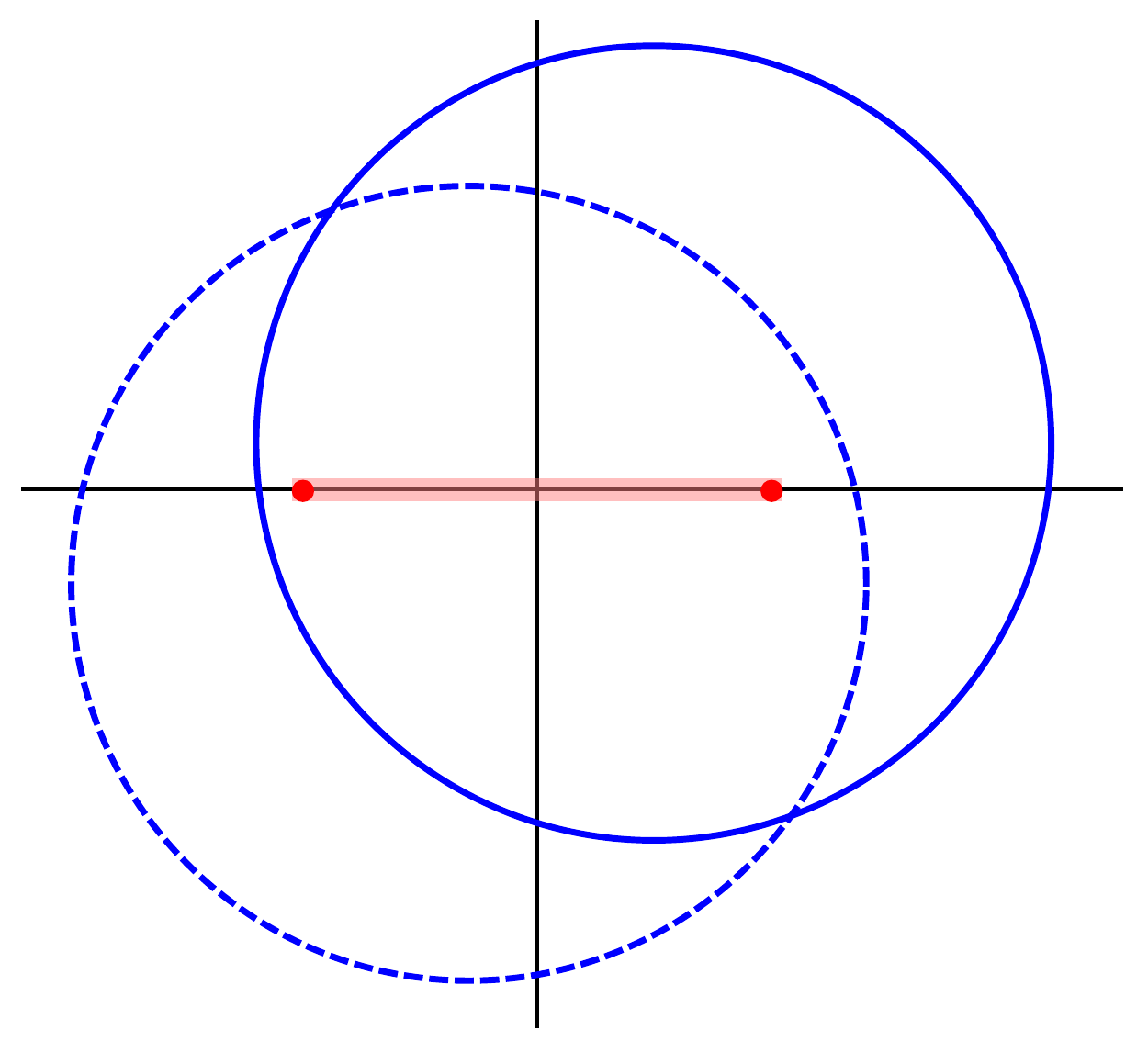}
\caption{For the entanglement entropy of an interval of length $L<R$, there is also a contribution from multi-walled instantons.
This example shows a configuration in the case $n=2$; the solid and dotted line instanton walls lie on different sheets.}
\label{fig:multi-center}
\end{center}
\end{figure}

Let us consider the contribution from integrating over a single one of these variables, $x$.
The integration region is
\be
\cR = \Bigl\{ |x-{\tfrac{L}{2}}|^2 + y^2 < R^2 \Bigr\} \cap \Bigl\{ |x+{\tfrac{L}{2}}|^2 + y^2 < R^2 \Bigr\}
\ee
whose area%
\footnote{The explicit value is
\be
A_{\cR_2} = \pi R^2 - \frac{L}{2}\sqrt{4R^2-L^2} - 2R^2\tan^{-1}\left(\frac{L}{\sqrt{4R^2-L^2}}\right) \,.
\ee
}
we denote $A_{\cR_2}$. Note that $A_{\cR_2} \to 0$ as $L\to 2R$.

We also require the functional determinant $\det'\cD$.
For this instanton, the slow modes factorize into $n$ copies of the slow modes for a single instanton in the flat ($n=1$) geometry, so that
\be
{\det}_s'\cD = ({\det}_s'\cD_1^{(1)})^n 
\,.
\ee
The fast modes are essentially as before, except that now we have two singularities in the interior region.
Hence 
\be
{\det}_f\cD = 
({\det}\cD_1^{(1)})^n 
\left(
    \frac{\det\tilde\cD_n}
    {(\det\tilde\cD_1)^{n}}
\right)^2
\,.
\ee
This gives the full determinant
\be
{\det}'\cD
= ({\det}'\cD_1^{(0)})^n \left(\frac{\det'\cD_1^{(1)}}{\det\cD_1^{(0)}}\right)^{2n}
\left(\frac{\det\tilde\cD_n}{(\det\tilde\cD_1)^n}\right)^2
\ee
For $L\gg m^{-1}$, the leading contribution to the 0-instanton partition function takes the form
\be
\cZ_n^{(0)} = 
\left[
    \frac{\det\cD_n^{(0)}}
        {(\det\cD_1^{(0)})^n}
\right]^{-1}
\Bigl[\det \cD_1^{(0)}\Bigr]^{-n/2}
\ee
allowing us to write
\begin{align}
Z^{(1)}_n(\cR_2) &= 
\frac{1}{2^n}
Z_n^{(0)}
A_\cR^n e^{-nS_0}
\left(\frac{S_0}{2\pi}\right)^n
\left[\frac{\det{}'\cD_1^{(1)}}{\det\cD_1^{(0)}}\right]^{-n/2}
\left[
\left(
    \frac
    {\det \cD_1^{(0)}}
    {\det \tilde\cD_1}
\right)^n
\frac{\det\tilde\cD_n}
    {\det \cD_n^{(0)}}
\right]^{-1} \\
&=
Z_n^{(0)}
(A_{\cR_2} e^{-S_0} K)^{n}
P_n^2 \,.
\end{align}
(Note that in this case, there is one negative eigenvalue per sheet, giving $2^{-n}$.)
The other contributions are
\begin{align}
\cZ_n^{(1)}(\cR_0) &= \cZ_n^{(0)}\cdot n A_{\cR_0}Ke^{-S_0} \\
\cZ_n^{(1)}(\cR_1) &= \cZ_n^{(0)}
A_{\cR_1}Ke^{-nS_0}P_n(-2\pi i\kappa)^{n-1}
\end{align}
with $A_{\cR_0}=2\pi R^2-2A_{\cR_2}$ and $A_{\cR_1}=A-A_{\cR_0}-A_{\cR_2}$.
As before, for $d=2$ only the instantons in region $\cR_0$ should be summed over in the dilute gas approximation, giving total partition function
\be
\cZ_n^{(1)}(L) =
\cZ_n^{(0)}e^{nA_{\cR_0}K e^{-S_0}}
\bigl(1 + A_{\cR_1}Ke^{-nS_0}P_n(-2\pi i\kappa)^{n-1}\bigr)
\bigl(1 + (A_{\cR_2}Ke^{-S_0})^nP_n^2\bigr)
\,.
\ee
The final contribution to the entanglement at leading order is hence
\begin{align}
\cS &= \cS^{(0)} 
+ A_{\cR_1}Ke^{-S_0}\bigl(
    1 - (\log P_n)'\vert_{n=1} - \log(-2\pi i\kappa e^{-S_0})
\bigr) \\
&\qquad
+ A_{\cR_2}Ke^{-S_0}\bigl(
    1 - 2(\log P_n)'\vert_{n=1} 
    - \log(A_{\cR_1}Ke^{-S_0})
\bigr)
\,.
\end{align}

The $L$-dependence of this expression is of some interest.
For $m^{-1}\ll L\ll 2R$, the entanglement entropy experiences steep growth in $L$ (due to the $\tan^{-1}$ appearing in $A_{\cR_2}$), while as $L\to 2R$ the result reduces to the one in~\req{eq:instanton entanglement}.
Therefore, instanton effects give rise to a functionally strong (though small in absolute terms) dependence on interval size up to the instanton scale, even when this scale is much larger than the correlation length.

\subsubsection*{Decay, thermalization, and volume contributions
\label{sec:thermalization}}
A system undergoing vacuum decay exhibits an energy gap between the false and true vacua.
After long times, the resulting energy should be contained in a roughly thermal soup of particles in a thermalized state.
The entanglement entropy of thermal states are well-known to have contributions scaling with volume. 
Yet our computations see only area law contributions; why should this be?

Our results are not in conflict with this situation due to the qualifier ``at late times''.
During the growth of a bubble, essentially all excess energy goes into accelerating the bubble wall~\cite{Coleman3}.
Particles are produced only after bubbles are sufficiently large and common that they begin to collide.
As the methods used here describe vacuum evolution only at small bubble density, we should not expect our computation to yield thermal behavior.
Our results should therefore be interpreted as the contribution to entanglement due to well-separated bubbles at short times.

\section{The theta vacuum and coherent sums over Fermi-surfaces}
\label{sec:theta vacuum}
We would finally like to turn to a somewhat more detailed study of theta vacua. In the examples considered above, there is no evidence that the true (theta) vacua of a gauge theory should exhibit exotic scaling of entanglement entropy. 
It was also argued in \cite{Swingle} that field theories belong to class $s=1$ in the s-source renormalization scheme, and in these models, the entanglement entropy of the ground state necessarily satisfies an area law. 
The ground state degeneracy of models in this class should not scale with system size.
The perturbative ground states of gauge theories, on the other hand, are infinite in number, being labeled by a winding number; imposing locality requires physical states to transform by a phase under shift in winding number, so that a physical vacuum is an infinite superposition of perturbative vacua. 
As we shall see, upon UV regularization this degeneracy can be understood as scaling with system size. In this case, the dilute gas approximation is no longer valid. Therefore it is worth inspecting theta vacua in greater detail.

\subsection{The Schwinger model}
The vacuum structure of a generic gauge theory would be quite complicated to analyze directly.
We therefore wish to focus on the simplest manifestation of $\theta$ vacua possible, and the (1+1)-dimensional Schwinger model \cite{Schwinger} provides an exactly solvable gem. 

The Schwinger model is nothing other than a (1+1)-d Dirac fermion minimally coupled to a gauge field, with action
\be
S_\text{Schwinger} = \int d^2x\, \bar{\psi} \gamma^\mu (i \partial_\mu - e A_\mu) \psi - \frac{1}{4}F^{\mu\nu}F_{\mu\nu}.
\ee
This theory is exactly solvable, since the Green's function of the fermion can be computed exactly for arbitrary $A$ \cite{Rothe}, allowing $\psi$ to be integrated out directly.

The theory can also be solved in the canonical formalism, and the ground state wavefunction has been explicitly written down in \cite{Iso, Azakov}.
Put the theory on a spatial circle of length $L$.
The theta vacuum takes the form (in temporal gauge $A_0 = 0$) 
\be
|\theta \rangle = \sum_{N=-\infty}^{\infty} \exp(i N \theta) |N, \text{gs} \rangle
\,. 
\ee 
Here, 
\be
\ket{N, \text{gs}} = f_0 (N, \lambda, c) (\prod_{n>0} U_n) |N\rangle,
\ee
where $c$ is the Wilson line $c=\frac{1}{L} \int_0^{L} dx\, A_1(x)$. 
The state $\ket{N}$ is essentially the Dirac sea, with the $N$ lowest-lying particles added to (or removed from) both of the positive- and negative-chirality sectors. 
(These numbers must coincide in a physical state, where $Q_\text{tot}=0$).
The unitary operators $U_n = \exp\{-\gamma_n (\tilde\jmath_{+,n}^\dag \tilde{\jmath}_{-,n}^\dag- \tilde{\jmath}_{+, n} \tilde{\jmath}_{-,n})\}$ are obtained from the currents $j_{\pm} = \bar{\psi}\gamma_{\pm}\psi$ by a Bogoliubov transformation:%
\footnote{The constants $\gamma_n$ are given by
\be
\cosh 2\gamma_n = \frac{1}{E_n}\left(
\frac{2\pi n}{L}+\frac{e^2L}{4\pi^2n}\right)
\qquad
\sinh 2\gamma_n = \frac{1}{E_n}\frac{e^2L}{4\pi^2n} .
\ee
Here $E_n=\sqrt{(2\pi n/L)^2 + e^2/\pi}$.
}
\be
\tilde{\jmath}_{n,\pm} = \cosh \gamma_n\, j_{n,\pm} + \sinh \gamma_n\, j^\dag_{n,\mp}.
\ee
The factor of $U_n$ thus arises from the Bogoliubov transformation diagonalizing the Hamiltonian. Since this Bogoliubov transformation defines the vacuum state of a massive bosonic mode, it can only lead to an area law scaling of the entanglement entropy. 
The function $f_0$ is given by
\be
f_0(N,\lambda, c) = \Bigl(\frac{ML}{\pi}\Bigr)^{1/4} e^{-( e c L - 2\pi N)^2/2\pi M L}
\ee
where $M = e/\sqrt{\pi}$.

What is noteworthy here is that the ground state wavefunction is an infinite sum of Fermi surfaces.
In a discretized setting, this sum would be bounded by the total number of fermion sites. 
This means that the number of states that the instanton hops between actually scales with system size. 
This is a rather peculiar feature, and it is tempting to suggest that it gives the wavefunction a chance of violating the area law more severely than a simple Fermi surface. 
In the following, therefore, rather than working with the exact ground state wavefunction of the Schwinger model, we study a toy version in which we dispose of the  oscillatory exponential factor and the gauge field, and focus on a coherent sum of Fermi surfaces. 
We would like to inspect how much extra entanglement such a sum can lead to.

\subsection{Sum over Fermi surfaces as a toy model}
\textbf{Setup.}\:
We work with a toy analog of the Schwinger model theta vacuum: a coherent superposition of Fermi surfaces. 
Our system is a $(1+1)$-dimensional lattice model with $L$ lattice sites. We make use of the Jordan-Wigner transformation mapping a 
chain of distinguishable spins to a system of spinless fermionic degrees of freedom. 
A Fermi surface is obtained by acting on the vacuum state $\ket{0}$ with fermionic creation operators,
\be
  \ket{\Psi_p}= \prod_{i=0}^{p-1} c_{k_{i}}^{\dagger} \ket{0}
  ,
\ee 
where 
\be
c_{k}=\frac{1}{\sqrt{L}}\sum_{j=0}^{L-1}c_{j} e^{-ikj}
\,,
\ee
$k_n=\frac{2\pi n}{L}$,
$c_{j}=\Big(\prod_{i<j}\sigma^{z}_{i}\Big)\sigma^{+}_{j}$, and $\sigma^{+}_{j}=(\sigma^{x}_{j}+i\, \sigma^{y}_{j})/2$. 
In this state, all momenta up to the Fermi level ($k_{F}=k_{p-1}$) are filled. 
To make contact with the theta vacuum, we now consider a linear superposition of Fermi surfaces:
\be
|\tilde \Psi\rangle =
\sum_{p=0}^{L-1}f(p) \ket{\Psi_p} = \sum_{p=0}^{L-1}f(p) \prod_{i=0}^{p-1}c_{k_{i}}^{\dagger} \ket{0}
\ee 
for some weight function $f(p)$.
We will explore the entanglement entropy for functional forms of $f(p)$ inspired by the Schwinger model.

\medskip\noindent
\textbf{Computing the entanglement entropy.}\: 
We first pick a contiguous subsystem of size $L_{A}\le L/2$, and compute the reduced density matrix $\rho_{A} = \Tr_{B} |\tilde \Psi\rangle\langle\tilde \Psi|$. 
$S_{EE}$ is given in terms of the eigenvalues of $\rho_A$.
We do this for several choices of $L$ and $L_A$, and compare the result with the case of a single Fermi surface. 
We find that the coherent sum weighted by $f(p)$ is characterized by more or less the same amount of entanglement as a single Fermi surface. It is possible to tune $f(p)$ to acquire extra entanglement, but the dependence on $L$ is nowhere as strong as a volume law.

\begin{figure}[h]
\begin{subfigure}[b]{0.25\textwidth}
    \includegraphics[height=2.5cm]{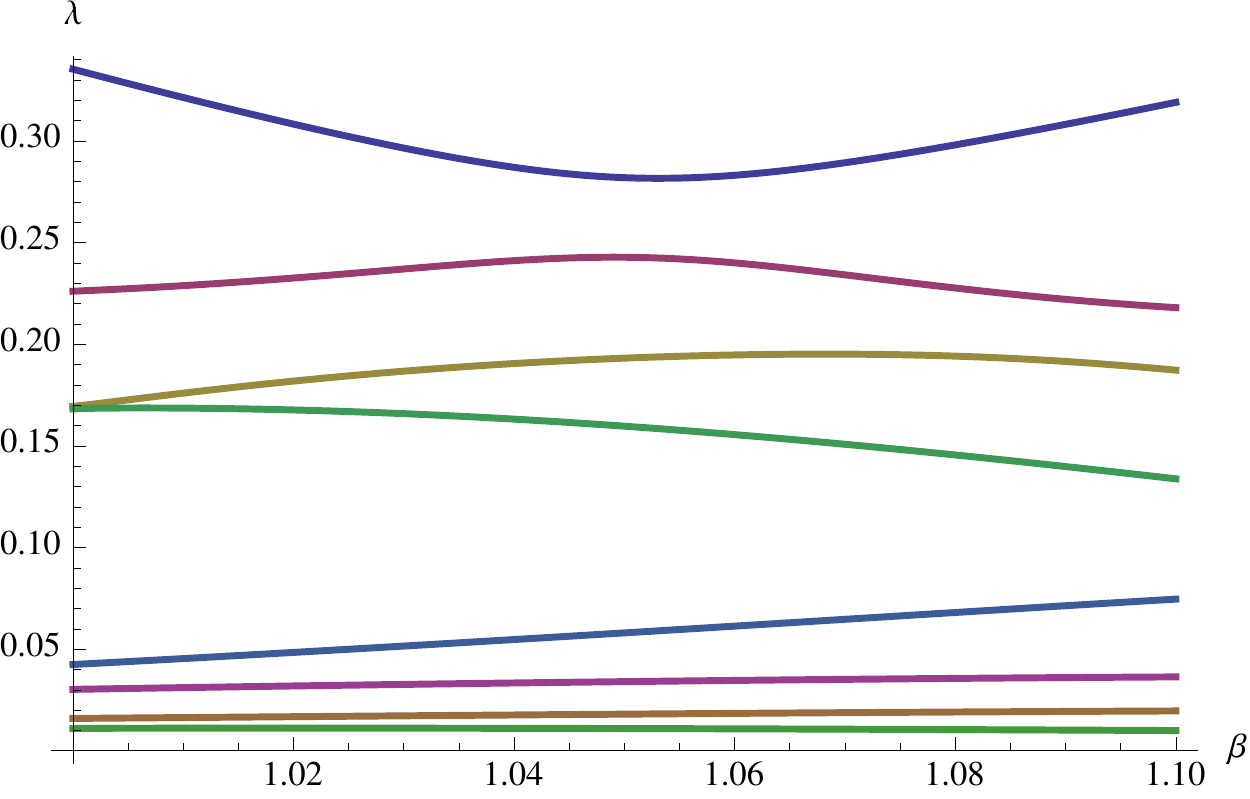}
\end{subfigure}%
\begin{subfigure}[b]{0.25\textwidth}
    \includegraphics[height=2.5cm]{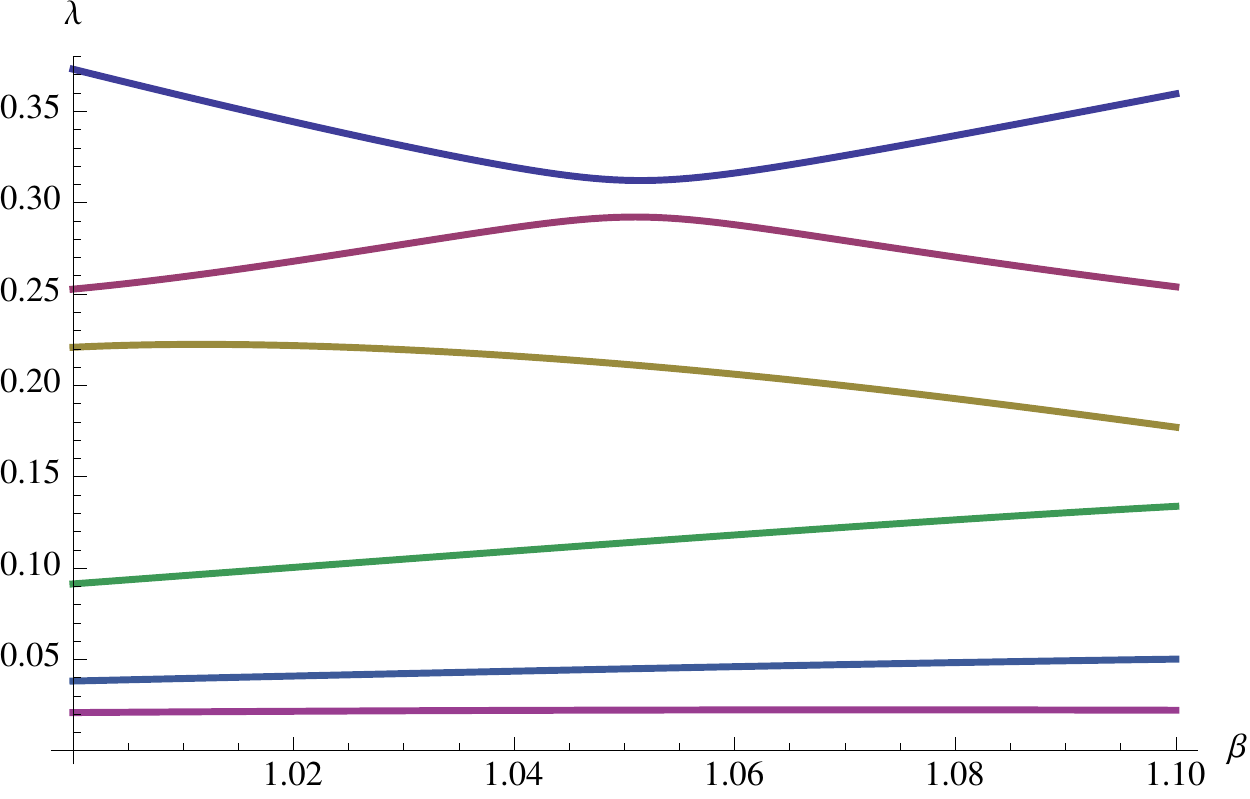}
\end{subfigure}%
\begin{subfigure}[b]{0.25\textwidth}
    \includegraphics[height=2.5cm]{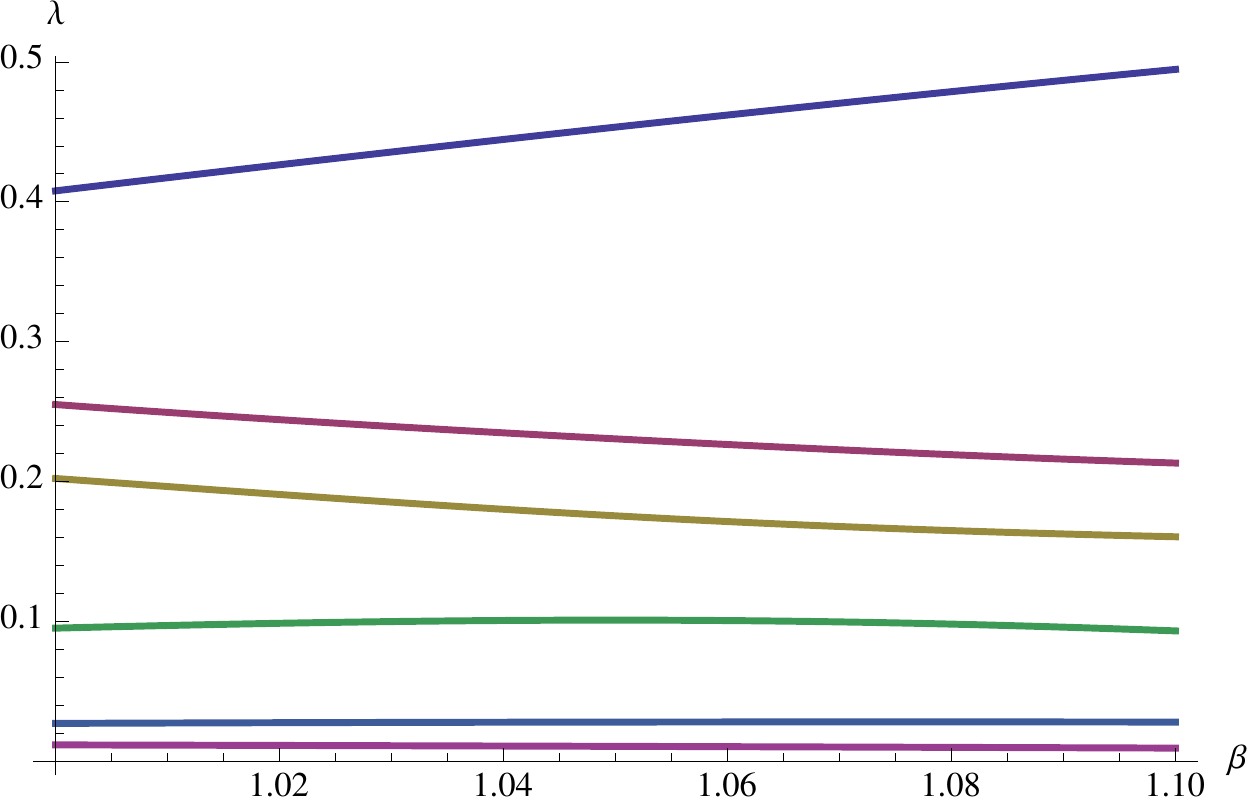}
\end{subfigure}%
\begin{subfigure}[b]{0.25\textwidth}
    \includegraphics[height=2.5cm]{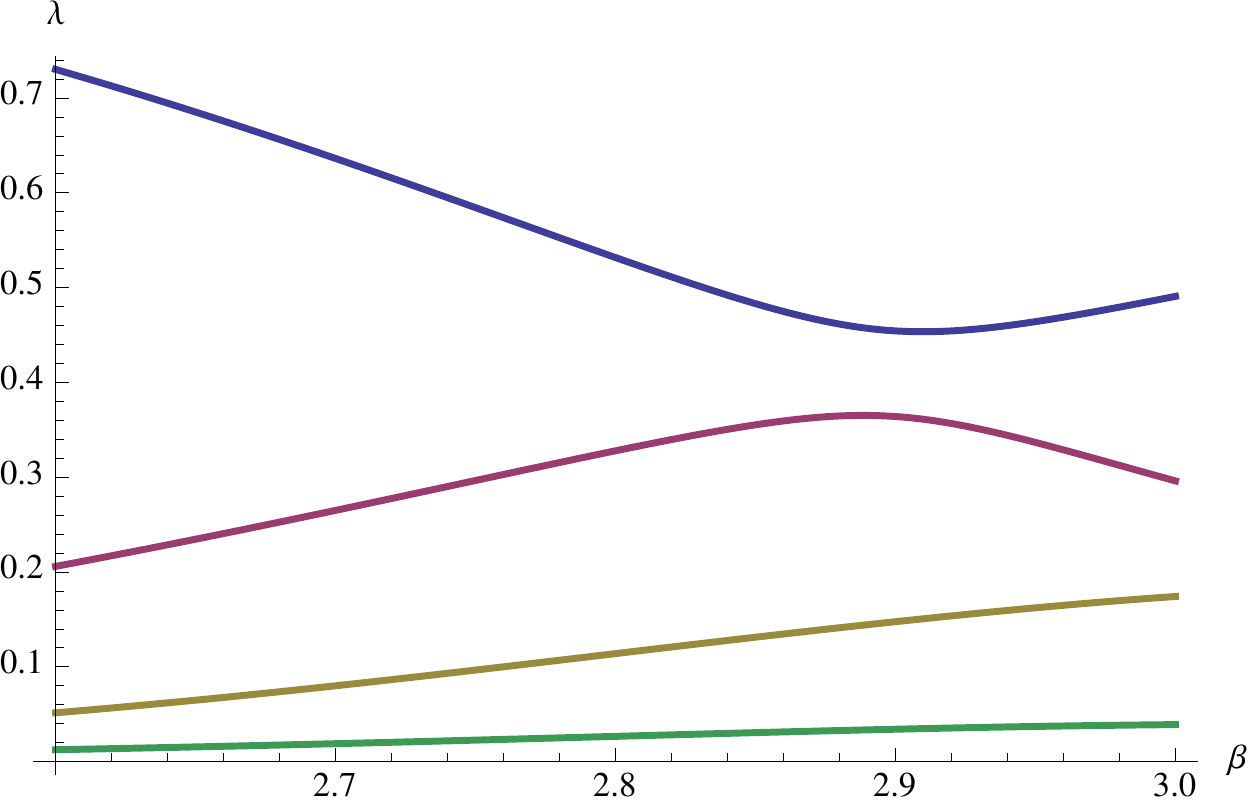} 
\end{subfigure}
\caption{Plots of the contributing eigenvalues of $\rho_A$ for superpositions of Fermi surfaces. The left-most figure corresponds to $L=8, L_{A}=4$, where 8 out of 16 eigenvalues contribute.
The second and third are for $L=8, L_{A}=3$ and $L=6, L_{A}=3$ respectively. In both of these cases 6 out of 8 eigenvalues contribute. The right-most plot is for $L=5, L_{A}=2$, where all 4 eigenvalues contribute.}
\label{fig:eigenvalues}
\end{figure}

\medskip\noindent
\textbf{Numerical results.}\: 
We first compute $S_{EE}$ of a single Fermi surface for several lattice sizes. 
We chose the values $L=4,6,7,8$; 
calculations up to $L=8$ are suffice to acquire a qualitative picture. 
For $L=8$ and $L_{A}=4$, if we consider a Fermi surface obtained by filling all modes up to $k=\frac{6 \pi}{L}$, we find from our numerical analysis that only 8 of the 16 eigenvalues of the reduced density matrix $\rho_{A}$ contribute to the entropy, as shown in the left-most plot of Figure \ref{fig:eigenvalues}. One can easily check that the number of contributing eigenvalues increases with system size. 
In 1d, where an ``area'' is a point, this constitutes a violation of the area law; this is expected in the presence of a Fermi surface. 

\begin{figure}[h]
\begin{subfigure}[b]{0.25\textwidth}
    \includegraphics[width=\linewidth]{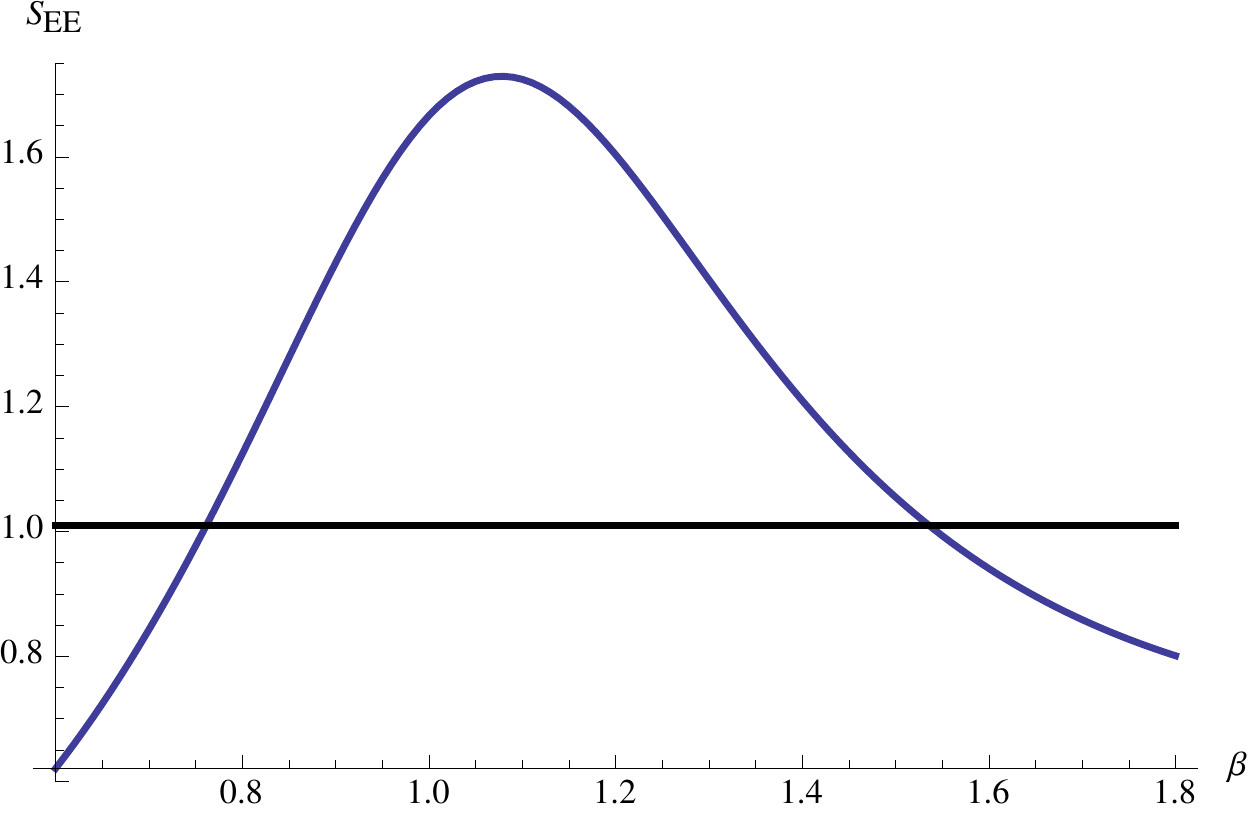}
    \caption{}
\end{subfigure}%
\begin{subfigure}[b]{0.25\textwidth}
    \includegraphics[width=\linewidth]{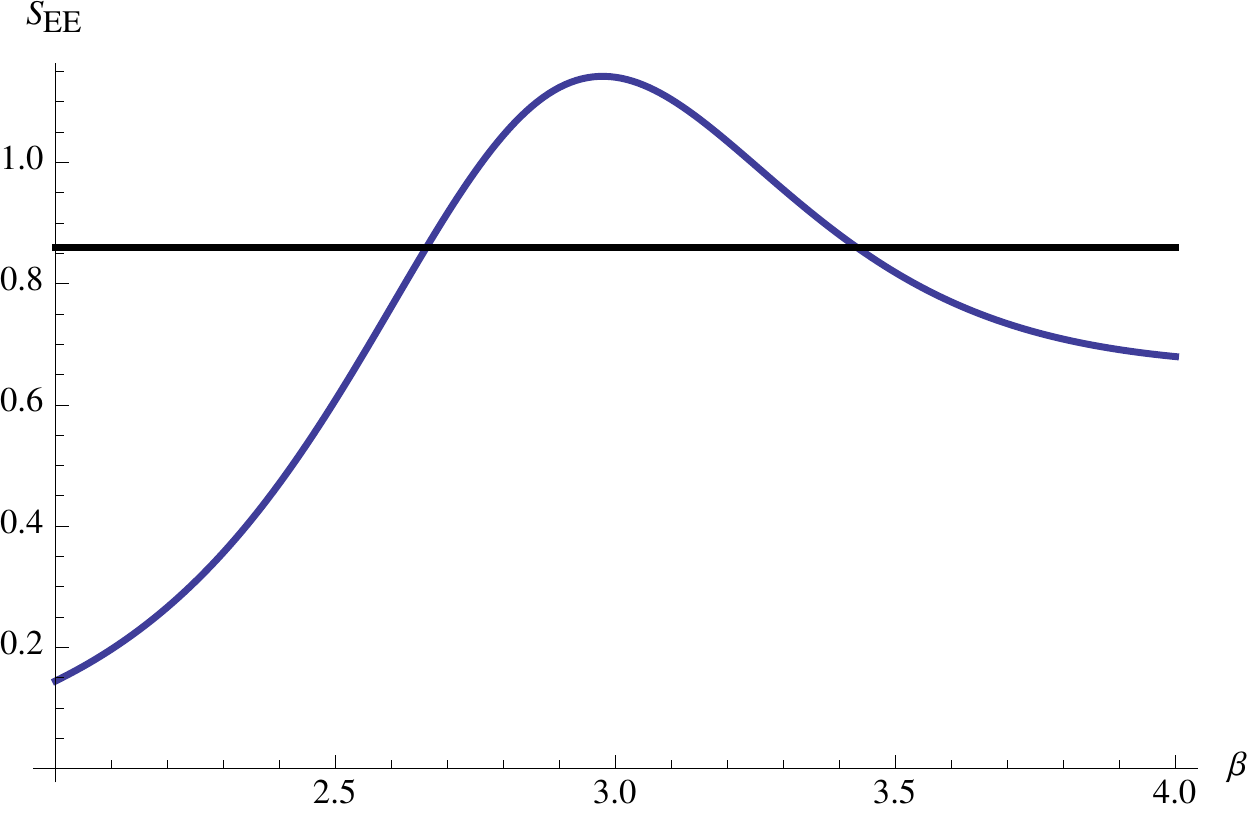}
    \caption{}
\end{subfigure}%
\begin{subfigure}[b]{0.25\textwidth}
    \includegraphics[width=\linewidth]{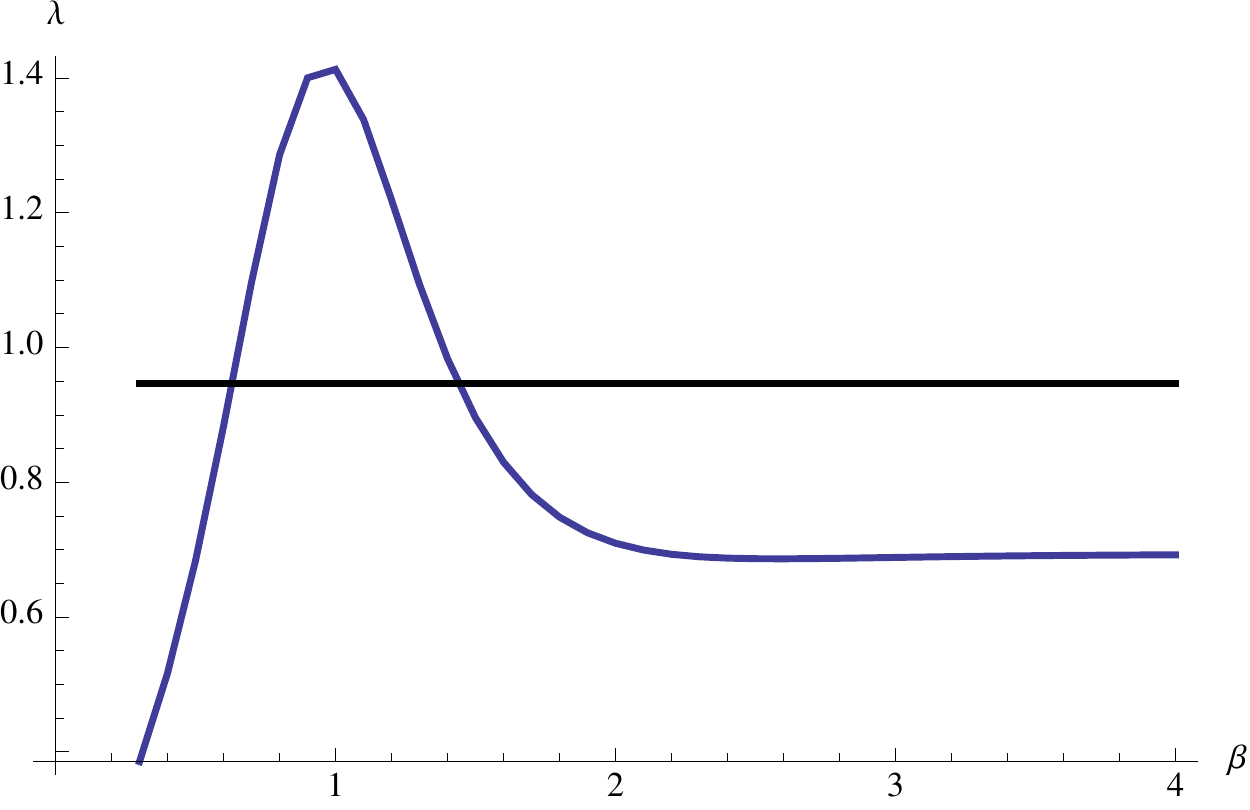}
    \caption{}
\end{subfigure}%
\begin{subfigure}[b]{0.25\textwidth}
    \includegraphics[width=\linewidth]{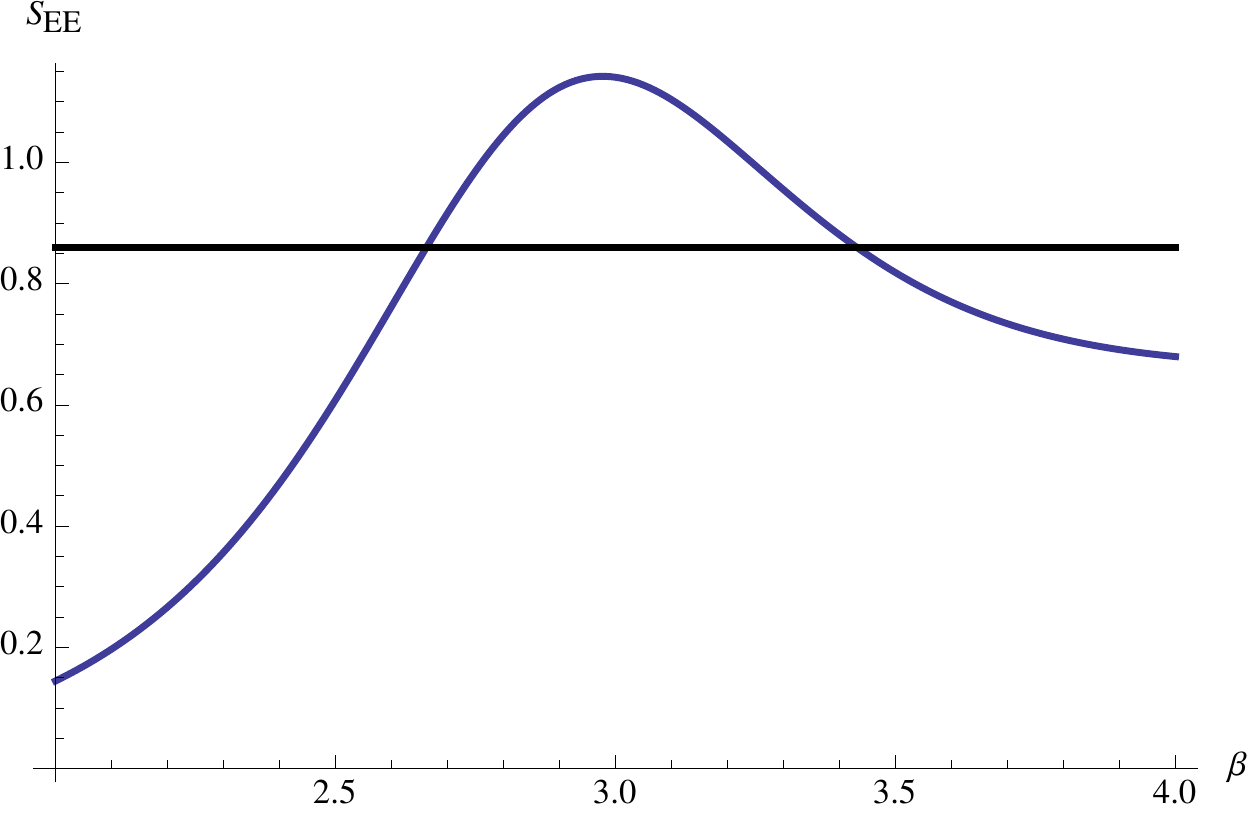}
    \caption{}
\end{subfigure}
\caption{Plots of $S_{EE}(\beta)$ for  superpositions of Fermi surfaces. 
The (sub-)system sizes are: (a) $L=8, L_{A}=4$, (b) $L=8, L_{A}=3$, (c) $L=6, L_{A}=3$, and (d) $L=5,L_{A}=2$.
For each of these cases, $S_{EE}$ of a single Fermi surface (obtained by filling all modes up to $k=\frac{6 \pi}{L}$) is shown in black.
Note that $S_{EE}$ is greater than that of a single Fermi surface only for a specific range of $\beta$. 
}
\label{fig:entropy}
\end{figure} 

Next, we consider the superposition of Fermi surfaces with weight functional of the form
\be
f(p)=\alpha e^{-p\,\beta}
\ee
in analogy to the theta vacuum of the Schwinger model. 
Figure~\ref{fig:entropy} gives plots of $S_{EE}(\beta)$ for four choices of (sub-)system size. 
The figures show that there is a mild enhancement in entanglement entropy for some range of $\beta$.
The $\beta$ dependence of the eigenvalues contributing to $S_{EE}$ is plotted in Figure~\ref{fig:eigenvalues} for the same four systems. 
These plots show that the entanglement entropy rises above that of a single Fermi surface precisely when more eigenvalues are contributing.
As we can see from the entanglement spectra, however, the number of contributing eigenvalues is not very sensitive to $\beta$. 
At most about 8 of the 16 eigenvalues contribute to the entropy --- more than for a single Fermi surface, but not by much. 

In all of the cases considered above, it is clear that entanglement entropy does not exhibit a volume law behavior, as this would require the number of participating eigenvalues to be on the order of $2^{L_A}$.

\section{Conclusions}
In this note, we initiated the study by instanton calculus of non-perturbative corrections to the entanglement entropy. 
Our computations were made by applying the replica trick directly to instanton solutions in the replica geometry.
Using the duality between the $\U(1)$ model and the XY-model, we showed that in this case our prescription preserves the duality map between path integral contributions, and therefore produces the correct entanglement entropy. 
Such insights allowed us to find explicit instanton solutions in the replica geometry, both in $\U(1)$ theories and in more general non-abelian Yang-Mills theories. 
Applying this prescription to several cases, we find that the non-perturbative contributions obey the area law. 
Moreover, for non-perturbative vacuum decay in $\phi^4$ theories in the dilute gas limit, area law behavior of the entanglement entropy can be demonstrated explicitly. 

We compared these results with numerical computations in discrete analogs of these models.
As an analog to the theta vacuum of the Schwinger model, we considered entanglement in a coherent superposition of Fermi surfaces. We also studied the time dependence of entanglement entropy in the perturbed transverse Ising model as an analog of non-perturbative vacuum decay in $\phi^4$ theory. 
In agreement with the instanton calculations, the area law is preserved in both cases.

\section*{Acknowledgements}
The authors would like to thank Chris Lau for contributing to the section on sums of Fermi-surfaces and collaboration on related problems. We would like to thank 
Horacio Casini and the referee for valuable comments.
AB, LYH and CMT would like to acknowledge the support of the Thousand Young Talents Program, and Fudan University. 
We would also like thank the organizers of the ``YKIS 2016: Quantum Matter, Spacetime and Information'' conference held at YITP, Kyoto in June 13-17, 2016, where part of this work was presented.

\appendix

\section{Corrections to the instanton solution}
In this appendix we comment on the perturbative correction to the Green's function in $2+1$ dimensions in the spirit of the section~\ref{Ob}. 
Before instantons are included, our theory is a massless scalar theory, so we can use conformal invariance to find the perturbative correction in $n$ to the Green's functions:
\be
\partial_n G_n|_{n=1} (x_1,x_2)=\int_{0}^{\infty} dx \int_{-\infty}^{\infty}dy \,\, x\, \corr{T_{\tau\tau}(0,x,y)\phi(x_{1})\phi(x_{2})}.
\ee
$x_{1}$ denotes the instanton center. 
The stress tensor is inserted at Euclidean time $\tau=0$, and without loss of generality we can set $x_{2}=(\tau_2,0,0)$. 

To proceed we require the three-point function $\corr{T_{\tau\tau}(0,x,y)\phi(\tau_1,x_1,y_1)\phi(\tau_2,0,0)}$. 
In three dimensions the conformal dimension of the scalar field is $\eta=\frac{1}{2}$. 
The correlator can be computed using the methods of~\cite{Osborn}:
\be
\corr{T_{\tau\tau}(0,x,y)\phi(x_{1})\phi(x_{2})}
=\frac{x_{23}^2}{x_{12}^3 x_{13}^2}t_{\tau\tau}( X_{23}),
\ee
where
\be
x_{23}=\sqrt{(\tau_1-\tau_2)^2+x_1^2+y_1^2},\;
x_{12}=\sqrt{\tau_1^2+(x-x_1)^2+(y-y_1)^2} ,\; x_{13}=\sqrt{(\tau_1-\tau_2)^2+x^2+y^2}
\nonumber
\ee
\be
t_{tt}( X_{23})= a\, h_{tt}(\hat X_{23})= (\hat X_{23})_{t}^2-\frac{1}{3},\;\;
(\hat X_{23})_{t}=\frac{(X_{23})_{t}}{\sqrt{X^2}},\;\;
(X)_{23}^2=\frac{x_{32}^2}{x_{13}^2x_{12}^2}.
\ee
The constant $a$ depends on the underlying theory.
In principle, one can plug all this back into the three-point function and perform the integration over $x$ and $y$ to find $\partial_{n}G_{n}|_{n=1}(x_1,x_2)$.


\end{document}